\documentclass{nature}
\usepackage[utf8]{inputenc}
\usepackage{color,ulem}
\usepackage{graphicx}
\usepackage{amssymb}
\usepackage{bm}
\usepackage{amsmath,hyperref,url}
\makeatletter
\let\saved@includegraphics\includegraphics
\AtBeginDocument{\let\includegraphics\saved@includegraphics}
\renewenvironment*{figure}{\@float{figure}}{\end@float}
\makeatother
\begin{document}

\title{A large-scale magnetic field produced by a solar-like dynamo in binary neutron star mergers}



\maketitle

\noindent
K. Kiuchi(Max Planck Institute for Gravitational Physics, CGQPI-YITP), A. Reboul-Salze(Max Planck Institute for Gravitational Physics), Masaru Shibata(Max Planck Institute for Gravitational Physics, CGPQI-YITP), Yuichiro Sekiguchi (Toho Univ.)
\vspace{1cm}


\begin{abstract}
The merger of neutron stars drives a relativistic jet, which must be driven by a strong large-scale magnetic field. However, the magnetohydrodynamical mechanism required to build up this magnetic field remains uncertain. By performing an ab-initio super-high resolution neutrino-radiation magnetohydrodynamics merger simulation in full general relativity, we show that the $\alpha\Omega$ dynamo mechanism, driven by the magnetorotational instability, builds up the large-scale magnetic field inside the long-lived binary neutron star merger remnant. As a result, the magnetic field induces the Poynting-flux dominated relativistic outflow with an isotropic equivalent luminosity $\sim 10^{52}$ erg/s and magnetically-driven post-merger mass ejection with the mass $\sim  0.1M_\odot$. Therefore, the magnetar hypothesis that an ultra-strongly magnetized neutron star drives a relativistic jet in binary neutron star mergers is possible. Magnetars can be the engines of short-hard gamma-ray bursts, and they should be associated with very bright kilonovae, which the current telescopes could observe. Therefore, this scenario is testable in future observation.  
\end{abstract}

The observation of GW170817/GRB 170817A/AT 2017gfo made binary neutron star mergers a leading player of multi-messenger astrophysics~\cite{LIGOScientific:2017vw,LIGOScientific:2017yn,Goldstein:2017mm,Savchenko:2017ff,Mooley:2018qf}. It revealed that at least a part of the origin of short-hard gamma-ray bursts and the heavy elements via the rapid-neutron capture process ($r$-process) is binary neutron star mergers~\cite{Goldstein:2017mm,LIGOScientific:2017yn,Savchenko:2017ff,Mooley:2018qf,Metzger:2010a,Lattimer:1974a,Eichler:1989a,Wanajo:2014wh}. 

All the fundamental interactions play an essential role in binary neutron star mergers. Thus, to theoretically explore them, a numerical-relativity simulation implementing all the effects of the fundamental interactions is the chosen way. Numerical relativity simulations manage to qualitatively explain AT 2017gfo, i.e., the kilonova emissions associated with the radioactive decay of the $r$-process elements~\cite{Shibata:2017xd,Kasen:2017sx,Metzger:2018qf,Fujibayashi:2017pu,Perego:2017wt,Waxman:2017sq,Kawaguchi:2018pt,Breschi:2021tb,Villar:2017wc}. 
However, a quantitative understanding of it is on its way. Moreover, there is no theoretical consensus about how the binary neutron star merger drove the short-hard gamma-ray burst~\cite{Ruiz:2017du,Mosta:2020hl,Fernandez:2018ka}. 

A relativistic jet was launched from the binary neutron star merger and was observed as a short-hard gamma-ray burst~\cite{Goldstein:2017mm,LIGOScientific:2017yn,Savchenko:2017ff,Mooley:2018qf}. Such relativistic jets are most likely driven by a magnetohydrodynamics process. This indicates that a binary neutron star merger remnant must build up a large-scale magnetic field via the dynamo to launch the jet~\cite{Brandenburg:2004j,Reboul-Salze:2021}. However, the mechanism of the large-scale dynamo in the merger remnant still needs to be clarified~\cite{Shibata:2021bb,Mosta:2015uc}. 

Also, the site for generating the large-scale magnetic field after the merger is a riddle. Is it inside a massive torus around a black hole formed after a merger remnant collapses into it? Or a long-lived remnant massive neutron star~\cite{Metzger:2010pp}? 
Recent numerical simulations for a black hole and torus system, as a remnant of binary neutron star mergers or black hole-neutron star mergers, suggest that a large-scale magnetic field is established after a long-term evolution of $O(1$--$10)~{\mathrm s}$~\cite{Hayashi:2022oxy,Gottlieb:2023est,Christie:2019lim,Fernandez:2018ka}. Consequently, a relativistic jet is launched via the Blandford-Znajek mechanism~\cite{Blandford:1977}. 
However, the long-lived remnant massive neutron star case is more computationally challenging than the black hole-torus system because the requirement to resolve relevant magnetohydrodynamical instabilities numerically is severe~\cite{Kiuchi:2017zz}. The physical mechanism for generating the large-scale magnetic field and jet launching could differ in the two scenarios mentioned above. 
It could be observationally testable if we managed to build a long-lived BNS model that generates a large-scale B-field and jet launching.

We tackle this problem with a super-high resolution neutrino-radiation magnetohydrodynamics simulation of a binary neutron star merger in full general relativity.

We employ our latest version of numerical-relativity neutrino-radiation magnetohydrodynamics code~\cite{Kiuchi:2022a}. We employ a static mesh refinement with 2:1 refinements to cover a wide dynamic range. For the simulations in this article, the grid structure consists of sixteen Cartesian domains, and each domain has a $2N \times 2N \times N$ grid in the $x,y$, and $z$-directions where we assume the orbital plane symmetry. We set $N=361$ and $\Delta x_{\rm finest}=12.5$~m. The employed grid resolution is the highest among the binary neutron star merger simulations~\cite{Kiuchi:2017zz}. 
The initial orbital separation is $\approx 44$~km (see Methods). 

We employ DD2~\cite{Hempel:2009mc} as an equation of state for the neutron star matter and symmetric binary with a total mass of $2.7M_\odot$. With this setup, a hypermassive neutron star transiently formed after the merger will survive for $>O(1)$~s~\cite{Fujibayashi:2020dv}. 

The purely poloidal magnetic-field loop is embedded inside the neutron stars with a maximum field strength of $B_{0,\mathrm{max}}=10^{15.5}$~G~\cite{Kiuchi:2015}. 
It is much stronger than those observed in binary pulsars of $10^7$--$10^{11}$\,G~\cite{Lorimer:2008se}. Available computational resources limit us to the choice of the strong field strength and idealized topology. Nonetheless, it is natural to anticipate that the magnetic-field amplification leads to a high field strength ($> 10^{16}$\,G) in a short timescale after the merge in reality (see below).

Figure~\ref{fig:Emag} plots the evolution of electromagnetic energy as a function of the post-merger time. As reported in Refs.~\cite{Kiuchi:2015,Kiuchi:2017zz}, the electromagnetic energy is exponentially amplified shortly after the merger due to the Kelvin-Helmholtz instability, which emerges at the contact interface when the binary merges. A part of the turbulent kinetic energy due to the instability is transferred to the electromagnetic energy~\cite{Kiuchi:2015}. In the inset, we generate the same plot for $-1 \le t - t_\mathrm{merger} \le 5$~ms but different initial magnetic-field strengths of $B_{0,\mathrm{max}}=10^{15}$\,G with red and $B_{0,\mathrm{max}}=10^{14}$\,G with orange while keeping the grid resolution. Also, we plot the results for the other grid resolutions of $\Delta x_{\rm finest}=100$~m and $200$~m for the brown and purple curves, respectively, while keeping $B_{0,\mathrm{max}}=10^{15.5}$\,G. It clearly shows that the growth rate of the electromagnetic energy for $0\lesssim t-t_\mathrm{merger}\lesssim 1$~ms does not depend on the employed initial field strength but on the employed grid resolution as expected from the properties of the Kelvin-Helmholtz instability~\cite{Price:2006,Rasio:1999} (see Extended Data Figure 1). 
The post-merger magnetic-field amplification due to this instability is terminated when the shear layer is dissipated by the shock waves at $t-t_\mathrm{merger}\approx 2$~ms.

At $t-t_\mathrm{merger} \approx 5$~ms, the electromagnetic energy temporarily settles to $\approx 3\times 10^{50}$~erg. However, the toroidal magnetic field (dotted curve) is subsequently amplified. In particular, its contribution to the total electromagnetic energy becomes prominent for $t-t_\mathrm{merger}\gtrsim 20$~ms and the growth rate is proportional to $t^2$, which indicates that an efficient magnetic winding occurs with a {\it coherent} poloidal magnetic field. 
The differential rotational energy of $\approx 1$--$2 \times 10^{53}~{\rm erg}$ of the remnant massive neutron star is the energy budget for the magnetic winding. Once the magnetic breaking starts to work, the electromagnetic energy saturates around $\approx 5 \times 10^{51}~{\rm erg}$. 
We also plot the result of the simulation with $\Delta x_{\rm finest}=200$~m in Fig.~\ref{fig:Emag}. 
With this resolution, it is hard to resolve the Kelvin-Helmholtz and magnetorotational instability, particularly in the high-density region (see Extended Data Figures 1--3, and Convergence study in Methods). 
Consequently, the electromagnetic energy amplification is less efficient in the low-resolution case. 
Particularly, there is a striking difference in the poloidal magnetic field between the simulations with $\Delta x_{\rm finest}=12.5$~m and $\Delta x_{\rm finest}=200$~m. 
Because the magnetic field amplified via the Kelvin-Helmholtz instability is randomly oriented~\cite{Kiuchi:2017zz}, there should be a mechanism to generate the coherent poloidal magnetic field.

To make it clear which part of the merger remnant is responsible for the generation of the coherent magnetic field, we evaluate the electromagnetic energy contained in the magnetorotational instability active region defined by $\rho \le 10^{14.5}~{\rm g/cm^3}$ in Fig.~\ref{fig:Emag_Mean_Field} (see Extended Data Figure 2). 
We also decompose the contributions from the mean-poloidal and toroidal magnetic fields by taking an axisymmetric average. 
It shows that the electromagnetic energy in this region is subdominant compared to that contained deep inside the core region with $\rho \ge 10^{14.5}~{\rm g/cm^3}$ (see also Fig.~\ref{fig:Emag}). However, it shows that the mean-poloidal field in the simulation with $\Delta x_\mathrm{finest}=12.5$~m exponentially grows from $20\lesssim t-t_\mathrm{merger}\lesssim 50~{\rm ms}$. At $t-t_\mathrm{merger}\approx 20~{\rm ms}$, its contribution to the total-poloidal field energy with the cyan-solid curve is $\sim 1\%$, and it becomes $\sim 10\%$ after the exponential growth. Such a clear exponential growth is invisible in the simulation with $\Delta x_\mathrm{finest}=200$~m for $t-t_\mathrm{merger}\lesssim 100~{\rm ms}$. Also, the mean-poloidal field energy differs by one or two order magnitudes in the  simulations with $\Delta x_\mathrm{finest}=12.5$~m and $200$~m. For the toroidal component, the mean- and total-field energy are in the same order for $t-t_\mathrm{merger}\gtrsim 30~{\rm ms}$. 
We also confirm the magnetoritaional instability is responsible for the mean poloidal-magnetic flux generation (see Extended Data Figures 4--5). 
It indicates a coherent magnetic-field generation is triggered via the magnetorotational instability. 

The $\alpha\Omega$ dynamo driven by the magnetorotational instability~\cite{Balbus-Hawley:1991} in the current context is a potential mechanism to generate the large-scale magnetic field~\cite{Brandenburg:2004j,Reboul-Salze:2021}. In the mean-field dynamo theory, we assume that the physical quantity $Q$ is composed of the mean field $\bar{Q}$ and the fluctuation $q$, i.e., $Q=\bar{Q}+q$. With it, we cast the induction equation into
\begin{align}
\partial_t \bar{\bf{B}}=\nabla \times \left(\bar{\bf{U}}\times \bar{\bf{B}}+\bar{\bf{\cal E}}\right), \label{eq:induction}
\end{align}
where $\bar{\bf{\cal E}}=\overline{\bf{u}\times\bf{b}}$ is the electromotive force due to the fluctuations, $\bf{B}=\bar{\bf{B}}+\bf{b}$ is the magnetic field, and $\bf{U}=\bar{\bf{U}}+\bf{u}$ is the velocity field. Note that the magnetorotational instability-driven turbulence produces the fluctuation $\bf{u}$ and $\bf{b}$. 
In the $\alpha\Omega$ dynamo, we express the electromotive force as a function of the mean magnetic field,
\begin{align}
\bar{\cal E}_i = \alpha_{ij} \bar{B}_j + \beta_{ij}\left(\overline{\nabla\times \bar{B}}\right)_j,\label{eq:emf}
\end{align}
where $\alpha_{ij}$ and $\beta_{ij}$ are tensors that do not depend on $\bar{B}_j$. We calculate the mean field by taking the average in the azimuthal direction. 

In the presence of a cylindrical differential rotation, the simplest mean field dynamo is an $\alpha \Omega$ dynamo, where the toroidal magnetic field is generated by the shear of the poloidal magnetic field by the differential rotation $\bar{U}_\phi$, also called $\Omega$ effect. The decreasing rotation with radius and Eq.~(\ref{eq:induction}) implies that the $\bar{B}_\phi$ should be anti-correlated with $\bar{B}_R$. To complete the dynamo cycle, the poloidal magnetic field has to be generated by the toroidal electromotive force $\bar{\cal E}_\phi$ with a main contribution from a diagonal component, $\alpha_{\phi\phi}$; the so-called $\alpha$ effect. $\bar{\cal E}_\phi$ is therefore correlated/anti-correlated to $\bar{B}_\phi$, depending on the sign of $\alpha_{\phi\phi}$ (Eq.~(\ref{eq:emf})). This complete cycle forms a dynamo wave that oscillates with a period of $P_\mathrm{theory} = 2\pi \lvert \frac{1}{2}\alpha_{\phi\phi}\frac{d\Omega}{d\ln R}k_z \rvert^{-1/2}$~\cite{Brandenburg:2004j,Reboul-Salze:2021} and propagates in the direction of $\alpha_{\phi\phi} \nabla \Omega \times e_\phi$, according to the Parker-Yoshimura rule~\cite{Parker:1955,Yoshimura:1975}. Here $k_z$ is the wave number of the dynamo wave in the vertical direction.
In this theoretical description, we have supposed that contributions from the other $\alpha_{ij}$ components and the turbulent resistivity tensor $\beta_{ij}$ are sub-dominant. We will show that it is the case in the following. 

First, we confirm that the employed simulation setup can capture the fastest-growing mode of the magnetorotational instability in the outer region of the hypermassive neutron star (see Extended Data Figure 2). Therefore, the turbulent state is developed.
Figure~\ref{fig:BF} plots the butterfly diagram for $\bar{B}_\phi$, $\bar{B}_R$, and $\bar{\cal E}_\phi$ at $R=30$~km. The top-left and right panels show the anti-correlation between $\bar{B}_\phi$ and $\bar{B}_R$, which indicates the $\Omega$ effect. To quantify the correlation between $\bar{B}_\phi$ and $\bar{\mathcal{E}}_\phi$, we compute the Pearson correlation $C_P(X,Y)$ between the two quantities, $X$ and $Y$, in the bottom-right panel of Fig.~\ref{fig:BF}. 
This figure shows that $\bar{B}_\phi$ and $\bar{\mathcal{E}}_\phi$ anti-correlate for $z\lesssim 15$~km, where the pressure scale height at $R=30$~km is $\approx 14.6$~km. 
The correlation between the electromotive force and mean current $\bar{J}_i=\left(\overline{\nabla \times \bar{B}}\right)_i$ is small, and thus, the $\beta_{ij}$ tensor can be neglected (see Extended Data Figure 6). 
Therefore, the generation of the mean poloidal magnetic field is determined primarily by the $\alpha_{ij}$ tensor. While $\alpha_{\phi R}$ has a non-negligible contribution, the contribution of $\alpha_{\phi\phi} B_\phi$ dominates in the turbulent region for $z\lesssim 10$ km (see Extended Data Figure 6). This shows that $\alpha_{\phi\phi}$ is the main component, which is plotted in the bottom-right panel of Fig.~\ref{fig:BF}. 
We also confirm the $\alpha^2\Omega$ dynamo is irrelevant compared to the  $\alpha\Omega$ dynamo (see Extended Data Figure 7). 

With these quantities, we can predict the period of the $\alpha\Omega$ dynamo listed in Table~\ref{tab:alpha_Omega}, where we measure the shear rate $q=-d\ln\Omega/d\ln R$ at the selected radius and choose the wave number $k_z$ corresponding to the pressure scale height, the most extended vertical length in the turbulent region. 
The sixth and seventh columns report the predicted period of the $\alpha\Omega$ dynamo and the period measured in the butterfly diagram. The agreement at $R=30$~km is remarkable. We also show the comparison at different radii from $R=20 $~km to $R=50$~km in the table, and the deal is also reasonable. 
In addition, since $\alpha_{\phi\phi}$ is negative and $q$ is positive, the dynamo wave propagates along the direction of the Parker-Yoshimura rule, i.e., the $z$-direction (\url{http://www2.yukawa.kyoto-u.ac.jp/~kenta.kiuchi/anime/SAKURA/Br_core_cut_9km.m4v}).
With these findings, we conclude that the dynamo in our simulation can be interpreted as an $\alpha\Omega$ dynamo, and it builds up the large-scale magnetic field in the remnant hypermassive neutron star (see Fig.~\ref{fig:Emag}). 

After the development of the coherent poloidal magnetic field due to the $\alpha\Omega$ dynamo and resultant efficient magnetic winding, a magnetic-tower outflow is launched toward the polar direction (\url{http://www2.yukawa.kyoto-u.ac.jp/~kenta.kiuchi/anime/SAKURA/DD2_135_135_Dynamo.mp4}, see Methods). 
The mean-poloidal magnetic flux is generated in the magnetorotational instability active region. The mean-magnetic field deep inside core $(\rho \ge 10^{14.5}~{\rm g/cm^3})$, which could be a relic of the initial field, stay buried below $r\lesssim 10~{\rm km}$ in the polar region through out the simulation (\url{http://www2.yukawa.kyoto-u.ac.jp/~kenta.kiuchi/anime/SAKURA/movie_Mean_Poloidal_Flux.mp4}, see Methods). It indicates that the large-scale field generated by the $\alpha\Omega$ dynamo is responsible for the jet launch. 
We confirm that the low-resolution simulation cannot capture the strong Poynting flux-dominated outflow launching and the enormous post-merger ejecta~\cite{Mosta:2020hl} (see also Extended Data Figure 8).

The luminosity of the Poynting flux defined by $L_\mathrm{Poy}=-\oint_{r \approx 500~\mathrm{km}}\sqrt{-g}\left({T^r}_{t}\right)_\mathrm{(EM)}d\Omega$, where $T^{\mu\nu}_\mathrm{(EM)}$ is the stress energy tensor for the electromagnetic field, is $\approx 10^{51}$~erg/s at the end of the simulation of $t-t_\mathrm{merger}\approx 150$~ms. 
The high Poynting flux is confined in the region with $\theta \lesssim 12^\circ$ where $\theta$ is a polar angle as plotted in the top-left panel of Fig.~\ref{fig:SGRB} (see also  Extended Data Figure 8). 
The luminosity and jet opening angle are broadly compatible with some of observed short-hard gamma-ray bursts~\cite{Fong:2015oh}. We estimate the terminal Lorentz factor $\Gamma_\infty$ by the magnetization parameter $\sigma_\mathrm{LC}=b^2/4\pi\rho c^2$ at the light cylinder radius of $r_\mathrm{LC}=c/\Omega \approx 40~{\rm km}\left(\Omega/7000~{\rm rad/s}\right)^{-1}$ contained in a region with $\theta<12^\circ$~\cite{Drehnkhahn:2002,Metzger:2008}. We found the baryon loading is still severe, e.g., $\dot{M}_\mathrm{outflow} \sim 10^{-2}M_\odot/\mathrm{s}$ in the polar region. Nonetheless, $\Gamma_\infty$ keeps increasing with time, but is fluctuating. It reaches $\approx 10$--$20$ at the end of the simulation. Therefore, if the magnetic reconnection efficiently dissipated the Poynting flux, the Lorentz factor of $\sim 10$ could be possible~\cite{Drehnkhahn:2002}. Also, the evacuation due to the strong Poynting flux in the polar region continues at the end of the simulation. Therefore, a higher $\Gamma_\infty$ could be possible after the long-term evolution.

The dynamical ejecta driven by the tidal force and the shock wave during the merger phase is $\approx 0.002M_\odot$~\cite{Fujibayashi:2020dv}. After the development of the coherent magnetic field, we observe a new component in the ejecta driven by the Lorentz force, i.e., the magnetically-driven wind~\cite{Blandford:1982di}. The mass of this component is $\approx 0.1M_\odot$ at the end of the simulation $t-t_\mathrm{merger}\approx 150$~ms. The electron fraction of the post-merger ejecta shows a peak around $\approx 0.2$ with an extension to $\sim 0.5$. The terminal velocity of the post-merger ejecta peaks around $\approx 0.08c$ where $c$ is the speed of light. Therefore, it could contribute to a bright kilonova emission via the synthesis of $r$-process heavy elements~\cite{Metzger:2010pp,Tanaka:2013ana}. 

We speculate that the high luminosity state and the post-merger mass ejection would continue for $O(1)$~s which corresponds to the neutrino cooling timescale~\cite{Fujibayashi:2020dv,Metzger:2008jt}. As reported in Refs.~\cite{Fujibayashi:2020dv,Hayashi:2022oxy}, the torus  
expands due to the angular momentum transport facilitated by the magnetorotational instability-driven turbulent viscosity after the neutrino cooling becomes inefficient. As a result, the funnel region above the remnant neutron star expands and the jet could no longer be collimated. A long-term simulation for $O(1)$~s of a remnant massive neutron star is future work to be pursued. 

Also, a simulation with initially weakly magnetized binary neutron stars is necessary to confirm the picture reported in this article because the saturated magneto-turbulent state and the generation timescale of the coherent poloidal magnetic field due to the magnetorotational instability could depend on the initial magnetic field strength and flux~\cite{Salvesen:2015tps}. The recent large-eddy simulations of magnetized binary neutron star merger may give a hint for this issue~\cite{Palenzuela:2022,Aguilera-Miret:2023qih}. 
Even if the initial magnetic field has a much weaker strength and/or different topology from the one we assume, the saturation strength and field profile due to the Kelvin-Helmholtz instability in the merger remnant are similar to what we found in this article. 
Also, how the mean-poloidal magnetic field is set in reality after the merger is an open problem. If the mean-poloidal magnetic field just after the merger is a relic of the pre-merger poloidal field, i.e., $10^{10}$--$10^{11}~{\rm G}$ at maximum, it may take $\ln(10^4)\times 0.02~{\rm s} \sim 0.2~{\rm s}$ 
to reach the saturation strength of $10^{14}$--$10^{15}$~G for the mean-poloidal field where we assume the mean-poloidal field is exponentially amplified with the period of the $\alpha\Omega$ dynamo~\cite{Brandenburg:2004j} (see Fig.~\ref{fig:Emag_Mean_Field} and Table~\ref{tab:alpha_Omega}). Consequently, the jet may be launched at $O(0.1)$~s after the merger in reality. However, the interior structure of the magnetic field in the pre-merger neutron stars is not well understood, and the magnetic reconnection of the fluctuated poloidal field generated by the Kelvin-Helmholtz instability may enhance the mean-poloidal field after the merger. 
They should be explored as a future task. 

To summarize, we tackled the large-scale magnetic-field generation in the long-lived binary neutron star merger remnant by the super-high resolution neutrino-radiation magnetohydrodynamics simulation in general relativity. The magnetorotational instability-driven $\alpha\Omega$ dynamo generates the large-scale magnetic field. As a result, the launch of the Poynting-flux-dominated relativistic outflow in the polar direction and an enormous amount of magnetically-driven wind are induced. 
Our simulation suggests that the magnetar engine generates short-hard gamma-ray bursts and bright kilonovae  emission, which could be observed in the near-future multi-messenger observations. 

\section*{Methods}

\subsubsection*{Numerical method}\label{appendix:method}
Our code implements the Baumgarte-Shapiro-Shibata-Nakamura-puncture formulation to solve Einstein's equation~\cite{Shibata:1995we,Baumgarte:1998,Baker:2005,Campanelli:2005}. The code also employs the Z4c prescription to suppress the constraint violation~\cite{Hilditch:2012}. The fourth-order accurate finite difference is used as a discretization scheme in space and time. The sixth-order Kreiss-Oliger dissipation is also employed. The HLLD Riemann solver~\cite{Mignone:2008} and the constrained transport scheme~\cite{Gardiner:2007} are employed to solve the equations of motion of the relativistic magnetohydrodynamic fluid. The neutrino-radiation transfer is solved by the gray M1+GR-Leakage scheme~\cite{Sekiguchi:2012} to take into account neutrino cooling and heating.

\subsubsection*{Initial data}
We employ quasi-equilibrium initial data of irrotational binary neutron stars 
in the neutrino-free beta equilibrium derived in a previous paper~\cite{Fujibayashi:2020dv} using the public spectral library LORENE~\cite{LORENE_SM}. The initial orbital angular velocity is $G m_0\Omega_0/c^3\approx 0.028$ with $m_0=2.7M_\odot$, and the residual 
orbital eccentricity is $\approx 10^{-3}$. For our high-resolution study, the data is remapped onto the computational domain described in the section of Grid setup. 

\subsubsection*{Grid setup}
We employ the static mesh refinement with 2:1 refinement, i.e., a grid resolution of a coarser domain is twice that of a finer domain. All the domains are composed of concentric Cartesian domains with a fixed grid number $N$. The grid number is $2N\times 2N\times N$ in the $x$, $y$, and $z$-directions where we assume the orbital plane symmetry. We employ the sixteen domains with $N=361$ and the finest grid resolution of $\Delta x_{\rm finest}=12.5$~m. The first three finest domains whose size are $[-4.5:4.5~{\rm km}]^2\times[0:4.5~{\rm km}]$, $[-9:9~{\rm km}]^2\times[0:9~{\rm km}]$, and $[-18:18~{\rm km}]^2\times[0:18~{\rm km}]$, respectively, are employed to resolve the Kelvin-Helmholtz instability, which emerges on a contact interface (shear layer) when the two neutron stars merge. The initial binary separation is $\approx 44$~km, and the coordinate radius of the neutron star is $10.9$~km. Thus, the fourth finest domain with $[-36:36~{\rm km}]^2\times[0:36~{\rm km}]$ covers the entire binary neutron star. 

We start a simulation with $\Delta x_{\rm finest}=12.5$~m and sixteen static mesh refinement domains. At $\approx 30$~ms after the merger, we remove the two finest domains with $\Delta x_{\rm finest}=12.5$~m and $\Delta x=25$~m and continue the simulation with $\Delta x_{\rm finest}=50$~m until $\approx 50$~ms after the merger. Then, we remove the domain with $\Delta x_{\rm finest}=50$~m and continue the simulation with $\Delta x_{\rm finest}=100$~m. The timing for removing the finer domains is determined by monitoring the magnetorotational instability quality factor so as not to go below the critical value of $10$ after the removal (see below). This strategy allows capturing the efficient magnetic-field amplification via the Kelvin-Helmholtz instability and resolving the magnetorotational instability inside the remnant while saving computational costs. 

To check the validity of the removing finer domain procedure, we continue the simulation with $\Delta x_\mathrm{finest}=12.5$~m up to $t-t_\mathrm{merger}\approx 33$~ms (see the blue dashed curve in Fig.~\ref{fig:Emag}). We observe a $\approx 10\%$ decrease in the poloidal electromagnetic energy in the simulation with $\Delta x_\mathrm{finest}=50$~m compared to that with $\Delta x_\mathrm{finest}=12.5$~m around this time. 
Nonetheless, the poloidal electromagnetic energy increases around $t-t_\mathrm{merger}\approx 40$~ms in the simulation with $\Delta x_\mathrm{finest}= 50$~m (the green dashed curve in Fig.~\ref{fig:Emag}) due to the $\alpha$ effect we discuss in the main text. Also, there is no significant decrease in the toroidal electromagnetic energy after removing the finer domains at $t-t_\mathrm{merger}\approx 30$~ms. Similarly, there is no visible deterioration at the second removal of the finer domain at 
$t-t_\mathrm{merger}\approx 50$~ms. 

For the convergence test, we also perform the simulations with $\Delta x_{\rm finest} =100(200)$~m and $N=361(185)$. The number of domains is thirteen.

\subsubsection*{Equation of state}
We extend an original DD2 equation of state to the low-density and temperature region with the Helmholtz equation of state~\cite{Timmes_Method}. Because any high-resolution shock-capturing scheme does not allow the vacuum state, we employ the atmospheric prescription outside the neutron stars. Specifically, we set the constant atmospheric density of $10^{3}~\mathrm{g/cm^3}$ inside $r\le L_\mathrm{atm}$, and assume the power-law profile with $\rho_\mathrm{atm}=\mathrm{max}[10^3(L_\mathrm{atm}/r)^3~\mathrm{g/cm^3},\rho_\mathrm{fl}]$ for $r> L_\mathrm{atm}$ where we set $L_\mathrm{atm}=36$~km. The employed Helmholtz equation of state determines the floor value of the rest-mass density, which is $\rho_\mathrm{fl}=0.167~\mathrm{g/cm^3}$. We also assume the constant atmospheric temperature of $10^{-3}$~MeV. 

\subsubsection*{Kelvin-Helmholtz instability}\label{appendix:KH}

The top panel of Extended Data Figure 1 is the same as the inset in Fig.~1 in the main article but with additional data. The blue, brown, and purple-solid curves plot the results with $\Delta x_{\rm finest}=12.5$~m, $100$~m, and $200$~m, respectively, with $B_{0,\mathrm{max}}=10^{15.5}$~G. The red- and orange-solid curves show the results with $\Delta x_{\rm finest}=12.5$~m with $B_{0,\mathrm{max}}=10^{15}$~G and $10^{14}$~G, respectively.
The cyan curve shows the result with $\Delta x_{\rm finest}=18.75$~m and $B_{0,\mathrm{max}}=10^{14}$~G. The blue- and red-dotted curves plot the result with $\Delta x_{\rm finest}=12.5$~m and $B_{0,\mathrm{max}}=10^{14}$~G magnified by the square of the ratio of the initial magnetic-field strength, i.e., $(10^{15.5}/10^{14})^2=10^3$ and $(10^{15}/10^{14})^2=10^2$, respectively. 

This figure suggests the following two points: The blue(red)-solid and -dotted curves overlap until the back reaction starts to activate, which happens when the electromagnetic energy reaches $\approx 3\times 10^{49}$~erg. This implies that the magnetic field is passive for $t-t_\mathrm{merger}\lesssim 1$~ms, i.e., the linear phase. The saturation of the electromagnetic energy via the Kelvin-Helmholtz instability is likely to be $\approx 1$--$3\times 10^{50}$~erg corresponding to $O(0.1)\%$ of the internal energy~\cite{Kiuchi:2017zz,Aguilera-Miret:2020dhz,Aguilera-Miret:2021fre,Aguilera-Miret:2023qih}. 

The second point is that the growth rate of the electromagnetic energy in the linear phase is determined by the employed grid resolution, not by the employed initial magnetic field strength. To quantify the growth rate dependence on the grid resolution and initial magnetic-field strength, we estimate the growth rate by fitting the electromagnetic energy as $E_\mathrm{mag}(t)=A\exp\left(\sigma_\mathrm{KH}(t-t_\mathrm{merger})\right)$ for $0 \lesssim t - t_\mathrm{merge} \lesssim 1$~ms, which corresponds to the linear phase. The bottom panel of Extended Data Figure 1 plots the estimated growth rate as a function of the initial magnetic-field strength. The symbols denote the employed grid resolutions. With $\Delta x_{\rm finest}=12.5$~m, the growth rate is $\approx 7~\mathrm{ms}^{-1}$, irrespective of the initial magnetic-field strength. The growth rate increases with the grid resolution. With $\Delta x_{\rm finest}=100$ and 200~m, the efficient amplification cannot be captured. All the properties are consisten

We estimate how small $\Delta x_\mathrm{finest}$ should be to simulate the amplification due to the Kelvin-Helmholtz instability starting from the "realistic" initial magnetic-field strength of $10^{11}$~G~\cite{Lorimer:2008se}. The saturation strength found in the simulation with $\Delta x_\mathrm{finest}=12.5$~m is likely to be $|\bar{B}|\approx 10^{16.5}$~G. We fit the growth rate as a function of the grid resolution from Extended Data Figure 1, and it is $\sigma_\mathrm{KH}(\mathrm{ms^{-1}})=90/\Delta x_\mathrm{finest}(\mathrm{m})$ where we assume the inverse proportionality originating from the property of the Kelvin-Helmholtz instability~\cite{Price:2006,Rasio:1999}.
Suppose that the initial magnetic-field strength of $10^{11}$~G and the lifetime of the shear layer of $2$~ms, we estimate the required growth rate is
\begin{align}
\sigma_\mathrm{KH}=\frac{1}{2~\rm ms}\ln\left(\frac{10^{16.5}~\rm G}{10^{11}~\rm G}\right)^2=12.7~{\rm ms^{-1}}. 
\end{align}
Therefore, the required grid resolution is
\begin{align}
\Delta x_\mathrm{finest} = \frac{90}{\sigma_\mathrm{KH}}\approx 7.1~{\rm m}.
\end{align}

\subsubsection*{Magnetorotational instability, neutrino viscosity, and neutrino drag}
To quantify how well the magnetorotational instability is resolved in our simulation, we estimate the rest-mass-density-conditioned magnetorotational instability quality factor defined by 
\begin{align}
\langle Q_\mathrm{MRI} \rangle_\rho \equiv \frac{\langle \lambda_\mathrm{MRI} \rangle_\rho}{\Delta x} = \frac{1}{\Delta x}\frac{\int_{\rho} \lambda_\mathrm{MRI} d^3x}{\int_\rho d^3x},
\end{align}
where $\lambda_\mathrm{MRI}=2 \pi B^z/\left(\sqrt{4\pi\rho}\Omega \right)$ is the fastest growing mode of ideal magnetorotational instability~\cite{Balbus:1991}. As the condition in terms of the rest-mass density, we define a remnant core and a remnant torus by fluid elements with $\rho \ge 10^{13}~{\rm g/cm^3}$ and $<10^{13}~{\rm g/cm^3}$, respectively. Furthermore, we exclude the core region above $10^{14.5}~{\rm g/cm^3}$ in the estimate of the quality factor because in such a region, the radial gradient of the angular velocity is positive, and it is not subject to magnetorotational instability~\cite{Balbus:1991} as plotted in the top panel of Extended Data Figure 2. Also, we introduce the cutoff density of $10^7~{\rm g/cm^3}$ for the torus to suppress the overestimation of the quality factor in the polar low-density region.
The middle and bottom panels of the figure show that the fastest growing mode of magnetorotational instability is well resolved both in the core and torus throughout the simulation. Consequently, magneto-turbulence is developed and sustained. However, the magnetorotational instability, particularly in the core throughout the entire simulation and in the torus for $t-t_\mathrm{merger} \lesssim 80$--$90$~ms, can not be resolved with $\Delta x_{\rm finest}=200$~m. Thus, the turbulence is not fully produced in such a low-resolution run. 

One caveat is that the neutrino viscosity and drag could affect the magnetorotational instability as a diffusive and damping process~\cite{Guilet:2016}. The former (later) becomes relevant when the neutrino mean free path is shorter (longer) than the wavelength of the magnetorotational instability. Given a profile of the merger remnant, such as the density $\rho$, angular velocity $\Omega$, temperature $T$, and hypothetical magnetic-field strength $B^z_\mathrm{hyp}$, we solve the two branches of the dispersion relations to quantify the neutrino viscosity and drag effects~\cite{Guilet:2016}: 
For the neutrino viscosity, 
\begin{align}
\left[\left(\tilde{\sigma}+\tilde{k}^2\tilde{\nu}_\nu\right)\tilde{\sigma}+\tilde{k}^2\right]^2+\tilde{\kappa}^2\left[\tilde{\sigma}^2+\tilde{k}^2\right]-4\tilde{k}^2=0, \label{eq:SM_dis1} 
\end{align}
and for the neutrino drag,
\begin{align}
\left[\left(\tilde{\sigma}+\tilde{\Gamma}_\nu\right)\tilde{\sigma}+\tilde{k}^2\right]^2+\tilde{\kappa}^2\left[\tilde{\sigma}^2+\tilde{k}^2\right]-4\tilde{k}^2=0, \label{eq:SM_dis2}
\end{align}
where $\tilde{\sigma}=\sigma/\Omega$, $\tilde{k}=kv_A/\Omega$, $\tilde{\kappa}^2=\kappa^2/\Omega^2$, $\tilde{\nu}_\nu=\nu_\nu\Omega/v_A^2$, and $\tilde{\Gamma}_\nu=\Gamma_\nu/\Omega$. $\sigma$ and $k$ are the growth rate and wave number of the unstable mode of the magneorotational instability. $v_A=B^z_\mathrm{hyp}/\sqrt{4\pi\rho}$ is the Alfv\'{e}n wave speed. $\kappa^2$ is the epicyclic frequency. $\nu_\nu$ and $\Gamma_\nu$ are the neutrino viscosity and drag damping rate, respectively. The reference~\cite{Guilet:2016} provides their fitting formulae as a function of the rest-mass density and temperature:
\begin{align}
&\nu_\nu = 1.2\times 10^{10}\left(\frac{\rho}{10^{13}~\mathrm{g/cm^3}}\right)^{-2}\left(\frac{T}{10~\rm{MeV}}\right)^2~\mathrm{cm^2/s},\\
&\Gamma_\nu=6\times 10^3\left(\frac{T}{10 \mathrm{MeV}}\right)^6~\mathrm{1/s}.
\end{align}
The neutrino mean free path $l_\nu$ is also fitted by
\begin{align}
l_\nu = 2\times 10^3 \left(\frac{\rho}{10^{13}~{\rm g/cm^3}}\right)^{-1}\left(\frac{T}{10~\rm{MeV}}\right)^{-2}~\rm{cm}. 
\end{align}

Extended Data Figure 3 plots the growth rate of the magnetorotational instability for a given remnant massive neutron star profile and the hypothetical value of $B^z_\mathrm{hyp}$. We take the profiles on the orbital plane at $t-t_\mathrm{merger}\approx 10$~ms. The purple curve denotes the boundary where the condition $\tilde{\nu}_\nu$ or $\tilde{\Gamma}_\nu \approx 1$ is met~\cite{Guilet:2016}. Inside the boundary, the neutrino viscosity or the neutrino drag significantly suppresses the growth rate. Outside it, the growth rate is essentially the same as the ideal magnetorotational instability. On top of it, we plot the azimuthally-averaged magnetic field strength $\bar{B}^z_\mathrm{sim}$ obtained in the simulation. Because of the efficient amplification via the Kelvin-Helmholtz instability just after the merger, the neutrino viscosity and drag effects are irrelevant in the entire region of the merger remnant.

\subsubsection*{Generation of mean poloidal-magnetic flux via magnetorotational instability}
To validate our interpretation that the $\alpha\Omega$ dynamo is responsible for the large-scale magnetic field generation, we evaluate the mean poloidal-magnetic flux on a certain sphere of radius $r$ by
\begin{align}
\Phi_{\bar{B}_r}(r) = 2 \int^{\pi/2}_0 \bar{B}_r r^2 \sin\theta d\theta,
\end{align}
where $\bar{B}_r$ denotes the radial mean field in the spherical-polar coordinate.

Extended Data Figure 4 plots the evolution of the mean-poloidal magnetic flux. We select $8$~km and $20$~km as representative radii for the magneto-rotational instability inactive and active region (see the top panel of Extended Data Figure 2). On the one hand, in the simulation with $\Delta x_\mathrm{finest}=12.5$~m, the mean poloidal-magnetic flux on the sphere with $r=8$~km exhibits intensive time variability for $t-t_\mathrm{merger}\lesssim 20$~ms reflecting the oscillations of the remnant core, i.e., compression and decompression. The poloidal flux gradually decreases, presumably due to the magnetic reconnection, which is imprinted as the decrease of the total poloidal-field energy for $10 \lesssim t-t_\mathrm{merger} \lesssim 20$~ms in Fig.~\ref{fig:Emag}. 
Nonetheless, it relaxes to a roughly constant value and stays there until $t-t_\mathrm{merger}\lesssim 120$~ms. 
It reflects that magnetorotational instability is inactive in this region, i.e., no generation of the mean poloidal-magnetic flux. 
On the other hand, it is evident that the mean poloidal flux on the sphere with $r=20$~km is generated around $t-t_\mathrm{merger}\approx 25$~ms in the high-resolution simulation. Such an efficient mean poloidal flux generation is not observed in the simulation with $\Delta x_\mathrm{finest}=200$~m for $t-t_\mathrm{merger}\lesssim 80$~ms as plotted with the green-dashed curve in the figure (but see below for slight generation of the mean poloidal-flux between $40 \lesssim t-t_\mathrm{merger} \lesssim 60$~ms). 
For $t-t_\mathrm{merger}\gtrsim 100$~ms, the mean poloidal-magnetic flux is generated even in the low-resolution simulation because the magnetorotational instability starts to be partially, not fully, resolved in the low-density region (see the middle and bottom panels of Extended Data Figure 2). 
Therefore, we conclude that the magnetorotational instability is responsible for the generation of the mean poloidal-magnetic flux. 

Extended Data Figure 5 plots a meridional profile of the mean radial magnetic field, $\bar{B}_r$, at selected time slices for the simulation with $\Delta x_\mathrm{finest}=12.5~(200)$~m in the left (right) column (see also 
the link for the visualisation:\url{http://www2.yukawa.kyoto-u.ac.jp/~kenta.kiuchi/anime/SAKURA/movie_Mean_Poloidal_Flux.mp4}). 
It is evident that the mean poloidal field is generated in the magnetorotational instability active region ($\rho \le 10^{14.5}~{\rm g/cm^3}$), particularly, in the low-latitude region with $0.24 \lesssim \theta \lesssim 0.5\pi$. Then, it propagates towards the high-latitude region and generates the jet. 
We also point out that the polar region at a radius of $9$-$10$ km has a weak magnetic field. This indicates that the poloidal field below this region does not contribute towards the launch of a jet and stays buried throughout the simulation. 

The low-resolution simulation also shows some amplification of the mean poloidal field in the low-density region with $\rho\le 10^{12}~{\rm g/cm^3}$, where the magnetorotational instability is resolved. 
(see the visualisations for the mean poloidal flux and magnetorotational instability quality factor in the simulation with $\Delta x_\text{finest}=200$~m, respectively:\url{http://www2.yukawa.kyoto-u.ac.jp/~kenta.kiuchi/anime/SAKURA/movie_Mean_Poloidal_Flux_Low.mp4} and  \url{http://www2.yukawa.kyoto-u.ac.jp/~kenta.kiuchi/anime/SAKURA/movie_Mean_MRI_Qfac_Low.mp4}). 
It shows a qualitatively similar behavior to the simulation with 
$\Delta x_\mathrm{finest}=12.5~{\rm m}$, but the efficiency for generating the mean poloidal magnetic field is much lower, which leads to a weaker Poynting-flux luminosity (see the Detailed property of the Poynting-flux dominated outflow and magnetically-driven post-merger ejecta section).

\subsubsection*{$\alpha\Omega$ dynamo}\label{appendAO}
To understand the dynamo process in the simulation, we use the framework of the mean field theory with an axisymmetric average. We consider an axisymmetric average because it corresponds to the symmetry of the differential rotation in the system. In the mean-field dynamo theory, we assume that the velocity field $\bf{U}$ and the magnetic field $\bf{B}$ are respectively composed of the mean velocity field $\bar{\bf{U}}$ and velocity fluctuations $\bf{u}$, i.e., $\bf{U}=\bar{\bf{U}}+\bf{u}$ and of the mean magnetic field $\bar{\bf{B}}$ and the magnetic fluctuations $\bf{b}$. With it, we average the induction equation, which gives Eq.~(\ref{eq:induction}), where $\bar{\bf{\cal E}}=\overline{\bf{u}\times\bf{b}}$ is the electromotive force due to the fluctuations. 
To close the system, the electromotive force is often expressed as a function of the mean magnetic field and its spatial derivatives as decribed by Eq.~(\ref{eq:emf}). $\alpha_{ij}$ and $\beta_{ij}$ are respectively the tensors expressing the contributions of the mean magnetic field $\bar{\bf{B}}$ and its derivatives $\bar{\bf{J}} = \nabla \times \bar{\bf{B}}$ to the electromotive force. These tensors should not depend on $\bar{\bf{B}}$. Firstly, we focus on how the mean poloidal field is generated and then on the toroidal field.

To show how the mean poloidal field is generated by the alpha or beta effect, we assume that the mean velocity field is composed by the rotation speed $\bar{\bf{U}}=R \Omega \vec{e}_\phi$ and project the averaged induction equation Eq.~(\ref{eq:induction}) in the radial direction in cylinder coordinates, which gives:
\begin{align}
    \partial_t \bar{B}_R= - \partial_z \left[ \left(\bar{\bf{U}}\times \bar{\bf{B}}\right)_\phi+\bar{\bf{\cal E}}_\phi\right] =  - \partial_z \bar{\cal E}_\phi 
    = -\partial_z \left(\alpha_{\phi j} B_j + \beta_{\phi j} \left(\nabla \times \bar{\bf{B}}\right)_j\right), \label{eq:ind_R}
\end{align}
Similarly, by projecting the averaged induction equation in the vertical direction, the generation of the mean vertical field $\bar{B}_z$ is due to the azimuthal component of the electromotive force $\bar{\cal E}_\phi$. 
For the generation of the mean toroidal field, the $\Omega$ effect (i.e. the winding of the magnetic field by differential rotation) is also important and must be compare to the contributions of both the radial and vertical components of the electromotive force $\bar{\cal E}_R$ and $\bar{\cal E}_z$. 
To estimate which of the mean magnetic field $\bar{B}_i$ or the derivatives of the mean magnetic field $\bar{J}_i$ contributes the most to the electromotive force $\bar{{\cal E}}_i$, we compute the correlations between these quantities. The correlations are computed according to the Pearson correlation coefficient between the quantities $\bar{\cal E}_i$ and $Y_j= \bar{B}_j$ or $\bar{J}_j$  with the following formula:
\begin{equation}
    \mathcal{C}_P(\bar{\cal E}_i,Y_j) = \frac{\int_t (\bar{\cal E}_i - \langle \bar{\cal E}_i \rangle_t) (Y_j - \langle Y_j \rangle_t) dt }{\sqrt{(\int_t (\bar{\cal E}_i - \langle \bar{\cal E}_i \rangle_t)^2 dt)}\sqrt{(\int_t (Y_j - \langle Y_j \rangle_t)^2 dt)}}
    \label{eq:correlation},
\end{equation}
where $\langle \cdot \rangle_t$ represents a time average.

In the following sections we present the complementary analysis of the other contributions to the generation of the mean magnetic field than the $\alpha_{\phi \phi}$-effect and $\Omega$-effect. 
We therefore show the other correlations between the electromotive force and the magnetic field and the estimated values of the tensor component.
For the values of the alpha tensor components, several methods can be used. The simplest one is to estimate from the correlations \cite{Reboul-Salze:2021} but this method assumes that one component is dominant. In order to take into account the off-diagonal contributions, we compute the values of the alpha tensor coefficients in this study by using the singular value decomposition method to perform the least-square fit of mean-current data and mean field \cite{Racine:2011}.
 
\subsubsection*{Generation of the mean poloidal field $\bar{B}_R$ }\label{appendAO1}

As shown in Eq.~(\ref{eq:ind_R}), the generation of the axisymmetric poloidal field is due to the curl of the toroidal component of the electromotive force in the averaged induction equation.
In this section, we show the correlations between the toroidal electromotive force $\bar{{\cal E}}_\phi$ and the magnetic field components $\bar{B}_j$ and the mean current $\bar{J}_j$ (Top and middle panels of Extended Data Figure 6). The low correlations with the mean current $\bar{J}_j$ shows that its contributions to $\bar{{\cal E}}_\phi$ (i.e. the $\beta_{ij}$ tensor) can be neglected.  In the same way, the vertical magnetic field contribution $\bar{B}_z$ can be neglected. Extended Data Figure 6 also shows the anti-correlation of $\bar{B}_\phi$ and $\bar{{\cal E}}_\phi$ and that the radial magnetic field $\bar{B}_R$ is correlated to $\bar{{\cal E}}_\phi$ as well. This can be explained as the radial field $\bar{B}_R$ is anti-correlated to the toroidal magnetic field $\bar{B}_\phi$ due to the $\Omega$-effect (see Fig.~\ref{fig:BF}). Since the mean toroidal field $\bar{B}_\phi$ is anti-correlated to the toroidal electromotive force $\bar{{\cal E}}_\phi$, the mean radial field $\bar{B}_R$ is also correlated to it. To confirm that the contribution of $\bar{B}_\phi$ dominates, we first compute the non-diagonal components of the alpha tensor (Bottom panel of Extended Data Figure 6). We then compare the contribution of $\bar{B}_\phi$ and $\bar{B}_R$, that is respectively the time-averaged values of $\alpha_{\phi\phi} \bar{B}_\phi$ and $\alpha_{\phi R} \bar{B}_R$ in the first $10$ km. We obtain 
\begin{equation}
    \frac{\langle \alpha_{\phi\phi}\bar{B}_{\phi}\rangle_t}{\langle \alpha_{\phi R}\bar{B}_{R} \rangle_t}=1.87. 
\end{equation}

\subsubsection*{Generation of the mean toroidal field}\label{appendAO2}

In the main text, we show that the $\Omega$-effect is important to the generation of the mean toroidal field. In this subsection, we check whether the contribution of the $\alpha$-effect via the poloidal components of the electromotive force $\bar{{\cal E}}_R$ and $\bar{\cal E}_z$ is significant, in which case the dynamo is called an $\alpha^2 \Omega$ dynamo instead of a $\alpha\Omega$ dynamo. First, we confirm that the correlations between the electromotive force and the mean current $\bar{J}_j$ (Right panels of Extended Data Figure 7) are lower than with the mean magnetic field $\bar{B}_j$ (Left panels of Extended Data Figure 7). The contributions from the mean current can therefore be neglected. For the mean magnetic field, some components, for example the mean toroidal field $\bar{B}_\phi$, are strongly correlated  with the radial component of the electromotive force $\bar{\cal E}_R$. This indicates that the $\alpha$-effect might be important to the generation of the mean toroidal field. To compare the strength of these two effects, $\alpha$-effect and $\Omega$-effect, we computed the corresponding $\alpha$ tensor components $\alpha_{R \ i}$ and $\alpha_{z \ i}$ and estimated the ratio of the two dynamo numbers $C_\alpha = \max(\lvert\alpha_{R i}\rvert,\lvert\alpha_{z i}\rvert)R/\eta$ for $i\in[R,\phi, z]$ and $C_\Omega = \Omega R^2/\eta$, where $\eta$ is the resistivity, in the turbulent region averaged for one scale height, which gives at $R = 30$ km 
\begin{equation}
    \frac{C_\Omega}{C_\alpha} = \frac{\Omega R}{\max(\lvert\alpha_{Ri}\rvert,\lvert\alpha_{zi}\rvert)} \approx 30.8.
\end{equation}
The $\alpha$-effect contribution towards the generation of the mean toroidal field can therefore reasonably be neglected as the $\Omega$-effect dominates the generation of the mean toroidal field. The dynamo in our simulation can therefore be interpreted as an $\alpha\Omega$ dynamo.

\subsubsection*{Detailed property of the Poynting-flux dominated outflow and magnetically-driven post-merger ejecta}\label{sec:Poynting_flux_SM}
The link (\url{http://www2.yukawa.kyoto-u.ac.jp/~kenta.kiuchi/anime/SAKURA/DD2_135_135_Dynamo.mp4}) is a visualization for the rest-mass density (top-left), the magnetic field strength (top-second from left), the magnetization parameter (top-second from right), the unboundedness defined by the Bernoulli criterion (top-right), the electron fraction (bottom-left), the temperature (bottom-second from left), the entropy per baryon (bottom-second from right), and the geodesic criterion (bottom-right) on a plane perpendicular to the orbital plane. 

The top panel of Extended Data Figure 8 plots the angular distribution of the luminosity of the Poynting flux. 
The angular distribution of the luminosity for the Poynting flux is calculated by 
\begin{align}
L_\mathrm{Poynting}(\theta)=-\int_{r \approx 500~{\rm km}} \alpha \psi^6 r^2 ({T^r}_t)_\mathrm{(EM)}d\phi,
\end{align}
where $\alpha$ and $\psi$ are the lapse function and the conformal factor, respectively. The high luminosity of $\approx 2$--$8\times 10^{52}$~erg/s/angle 
is confined in a region with $\theta < 12^\circ$.

The middle panel plots the jet-opening-angle-corrected luminosity and luminosity of the Poynting flux with green and blue as functions of the post-merger time. According to Refs.~\cite{Meier:1999,Shibata:2011,Kiuchi:2012}, the luminosity of the Poynting flux-dominated outflow driven by the efficient magnetic winding from the binary neutron star merger remnant is estimated by
\begin{align}
L_\mathrm{Poy}\sim 10^{51}~{\rm erg/s} \left(\frac{\bar{B}_P}{10^{15}~{\rm G}}\right)^2
\left(\frac{R}{10^6~{\rm cm}}\right)^3\left(\frac{\Omega}{8000~{\rm rad/s}}\right),
\end{align}
where $\bar{B}_P$ is the azimuthally-averaged poloidal magnetic field. In the current simulation, it is $\sim 10^{15}~{\rm G}$. Thus, the Poynting flux luminosity found in the simulation is consistent with this estimation. 
The jet-opening-angle corrected luminosity is $\sim 10^{52}$~erg/s. Thus, if we assume the conversion efficiency to gamma-ray photons is $\sim 10\%$, it is compatible with the observed luminosity of short-hard gamma-ray bursts~\cite{Fong:2015oh}. 

The bottom panel plots the eject as a function of the post-merger time. The solid curve denotes the result by the Bernoulli criteria. The colored region in the inset shows the error of the baryon mass conservation, and it is below $10^{-7}M_\odot$ throughout the simulation.

\subsubsection*{Convergence study and effect of initial large-scale magnetic field}
Because our choice of the initial large-scale magnetic-field strength is much stronger than those observed in binary pulsars~\cite{Lorimer:2008se}, 
we should clarify whether such a strong large-scale field is responsible for the jet launching.

We begin by estimating the magnetic winding timescale originating from an initial magnetic field. Suppose we consider a binary neutron star merger with the highly magnetized end of $10^{11}$~G in the observed pulsars. In that case, the magnetic winding timescale to reach the saturation is
\begin{align}
\bar{B}_\phi \sim 10^{16.5}~{\rm G}\left(\frac{\bar{B}_R}{5\times10^{11}~{\rm G}}\right)\left(\frac{\Omega}{8000~{\rm rad/s}}\right)\left(\frac{t}{8~{\rm s}}\right), \label{eq:SM_winding}
\end{align}
where we assume the compression at the merger amplifies the initial poloidal field by a factor of five, which should be proportional to $\rho^{2/3}$ because of the magnetic-flux conservation.
We also assume the saturation field strength of $10^{16.5}~{\rm G}$ as suggested in the super-high resolution simulation (see Extended Data Figure 1). 
Therefore, the magnetic winding originating from the initial magnetic field should be minor or irrelevant in reality.

However, the magnetic winding timescale is much shorter than those in reality in both simulations with $\Delta x_\text{finest}=12.5$~m and $200$~m because of the assumed initial strong field.
The low-resolution is fine enough to resolve the compression and winding but can only partially capture the Kelvin-Helmholtz and the magnetorotational instability. 
Therefore, there would be no striking difference between the simulations for the jet launching mechanism if such a strong initial poloidal field and subsequent magnetic winding were responsible for it. 
The middle panel of Extended Data Figure 8 shows that the Poyinting-flux dominated outflow is launched at $t-t_\text{merger}\approx 60~{\rm ms}$ and reaches up to $\sim 10^{49-50}{\rm erg/s}$ in the low-resolution simulation with $\Delta x_\text{finest}=200~{\rm m}$. However, in the same plot, the super-high resolution simulation with $\Delta x_\text{finest}=12.5~{\rm m}$ shows that the Poynting-flux dominated outflow is launched at $t-t_\text{merger}\approx 35$~ms. The luminosity reaches up to $\sim 10^{51}{\rm erg/s}$. 
The difference is striking. As discussed in Extended Data Figure 5, the efficiency of the generation of the mean poloidal-flux in the super-high resolution simulation is much higher that in the low-resolution simulation. It indicates the efficient $\alpha\Omega$ dynamo is responsible for the strong jet launching.

The bottom panel of Extended Data Figure 8 also suggests that the Lorentz force-driven post-merger ejecta found in the run with $\Delta x_{\rm finest}=200$~m is launched. 
However, the amount of the ejecta mass is about one order of magnitude smaller than those in the run with $\Delta x_\text{finest}=12.5$~m, showing that the enhanced activity of magnetohydrodynamics effects by the efficient dynamo action plays an important role in ejecting matter~\cite{Mosta:2020hl}.

\section*{Data Availability}
The simulation raw data (265TB) that support the findings of this study are available from the corresponding author upon request. The prerequisite for the data request is to prepare a server to transfer it. 

\section*{Code Availability}
The numerical relativity code and data analysis tool are available from the correspond author upon reasonable request. 

\section*{Acknowledgements}
This work used computational resources of the supercomputer Sakura clusters at the Max Planck Computing and Data Facility. 
The simulation was also performed on Cobra, Raven at the Max Planck Computing and Data Facility, Fugaku provided by RIKEN, and the Cray XC50 at CfCA of the National Astronomical Observatory of Japan. This work was in part supported by the Grant-in-Aid for Scientific Research (grant Nos. 23H01172 for K.K, 20H00158 and 23H04900 for M.S. and Y.S.) of Japan MEXT/JSPS, and by the HPCI System Research Project (Project ID: hp220174, hp230204, hp230084 for K.K., M.S., and Y.S.). 
K.K. thanks to the Computational Relativistic Astrophysics members in AEI for a stimulating discussion. 
K.K. also thanks to Kota Hayashi and Koutarou Kyutoku for the helpful discussion and providing the initial data.

\section*{Author Contributions Statement}
K.K. and A.R.S. were the primary drivers of the project and wrote the main text and method sections and developed the figures. M. S., Y. S., and K.K. are primarily developers of the numerical relativity code from scratch. K.K. performed the numerical relativity simulations. A.R.S. analyzed the simulation data. 
M. S. prepared an infrastructure of large-scale numerical computations. 
All authors were involved in interpreting the results and discussed the results, and commented on and/or edit the text.

\section*{Competing Interests Statement} 
The authors declare no competing interests.

\section*{Tables}

\begin{table}
\begin{flushleft}
\caption{\bf The $\alpha\Omega$ dynamo period prediction and simulation data at several radii}\label{tab:alpha_Omega}
\begin{tabular}{lllllll}
\hline 
$R$ (km) & $\alpha_{\phi\phi}$ (cm/s) & $\Omega$ (rad/s)  & Shear rate & $k_z$ (/cm) &  $P_\mathrm{theory}$ (s)  & $P_\mathrm{sim}$ (s) \\
\hline 
$20$    & $-8.1\times 10^6$   & $4025$  & $q=1.00$ & $6.3\times 10^{-6}$ & $0.020$ & $0.018$   \\ 
$30$    & $-1.0\times 10^7$   & $2515$  & $q=1.34$ & $4.2\times 10^{-6}$ & $0.021$ & $0.018$--$0.024$   \\
$40$    & $-1.0\times 10^7$   & $1688$  & $q=1.44$ & $3.3\times 10^{-6}$ & $0.037$ & $0.018$--$0.030$   \\ 
$50$    & $-4.4\times 10^6$   & $1200$  & $q=1.50$ & $2.6\times 10^{-6}$ & $0.062$ & $0.030$--$0.040$\\
\hline 
\end{tabular}
$P_\mathrm{theory}$: The $\alpha\Omega$ dynamo period, $P_\mathrm{sim}$: Butterfly diagram period in the simulation 
\end{flushleft}
\end{table}

\section*{Figure Legends/Captions}

\clearpage

\begin{figure}[t]
\centering
\includegraphics[width=0.6\linewidth]{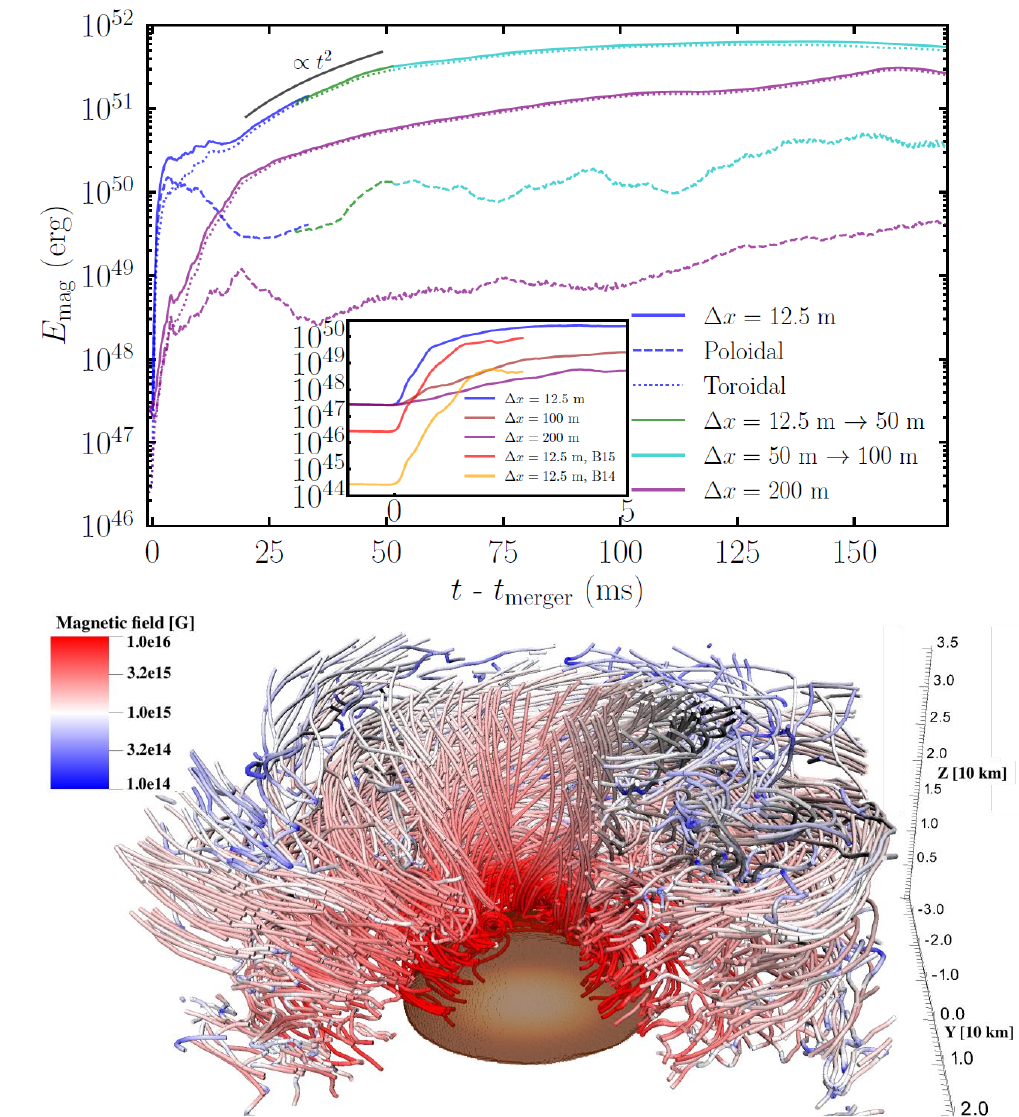}
\caption{\textbf{Overview of the magnetic field evolution in the binary neutron star merger remnant}. (Top) Electromagnetic energy as a function of the post-merger time for the total (solid), the poloidal (dashed), and the toroidal (dotted) components. The purple curves are for the simulation with $\Delta x_{\rm finest}=200$~m. 
The inset shows how the magnetic field amplification due to the Kelvin-Helmholtz instability depends on the initial magnetic-field strength and employed grid resolution. 
(Bottom) Magnetic field lines for the density of $\rho < 10^{13}$\,g\,cm$^{-3}$ at $t-t_\mathrm{merger}\approx 130$~ms. The core of the hypermassive neutron star is shown for the density of $\rho > 10^{13}$\,g\,cm$^{-3}$.
}\label{fig:Emag}
\end{figure}

\begin{figure}[t]
\centering
\includegraphics[width=0.6\linewidth]{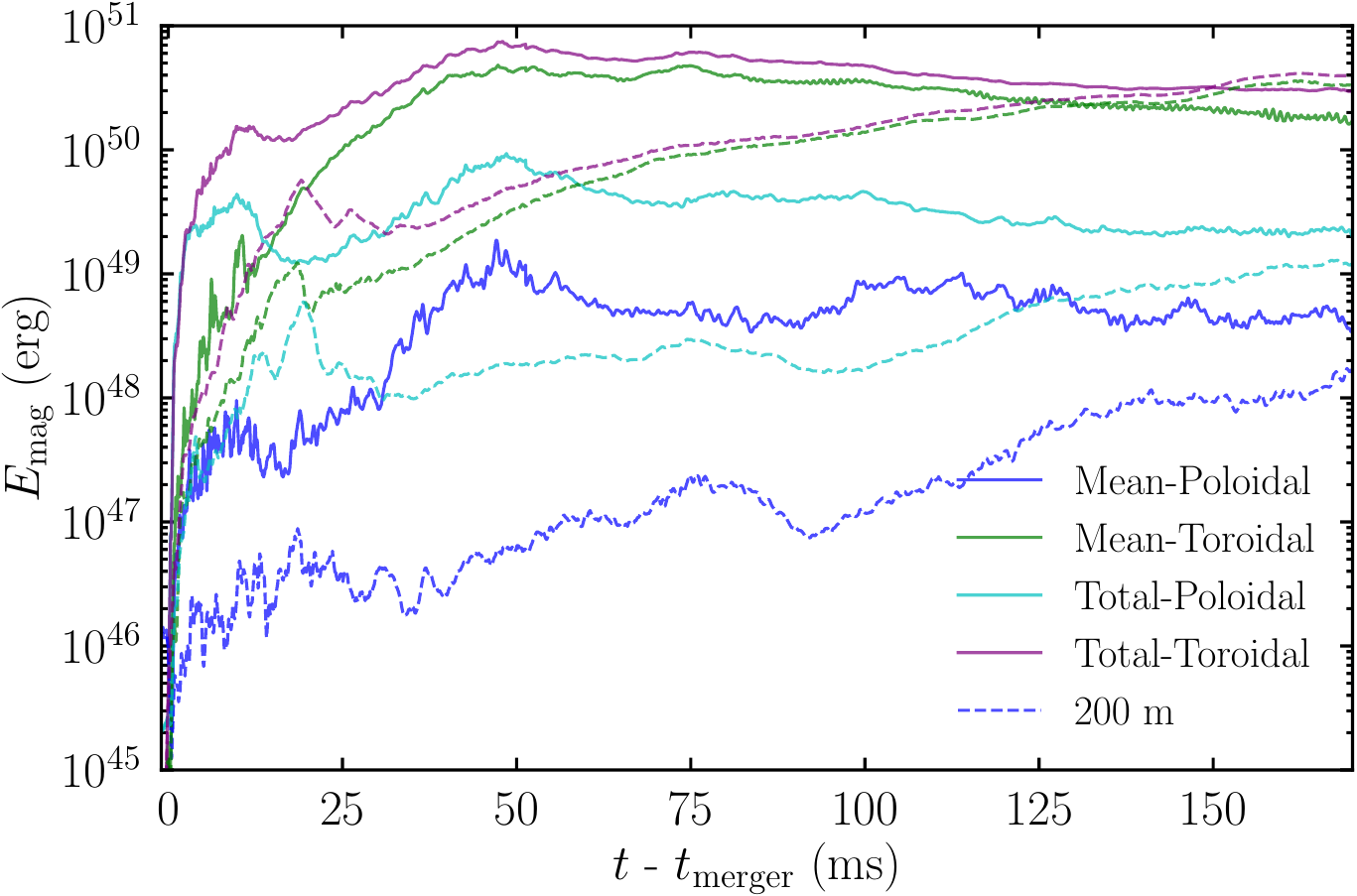}
\caption{\textbf{Generation of mean electromagnetic field}. Mean and total electromagnetic energy in magnetorotational instability active region with $\rho \le 10^{14.5}~{\rm g/cm^3}$ as a function of the post merger time. The blue, green, cyan, and purple curves denote the mean-poloidal, mean-toroidal, total-poloidal and total-toroidal component, respectively, in the high resolution simulation. The dashed curves are for the simulation with $\Delta x_\mathrm{finest}=200$~m.
}\label{fig:Emag_Mean_Field}
\end{figure}

\begin{figure}
\centering
\includegraphics[width=1.0\textwidth]{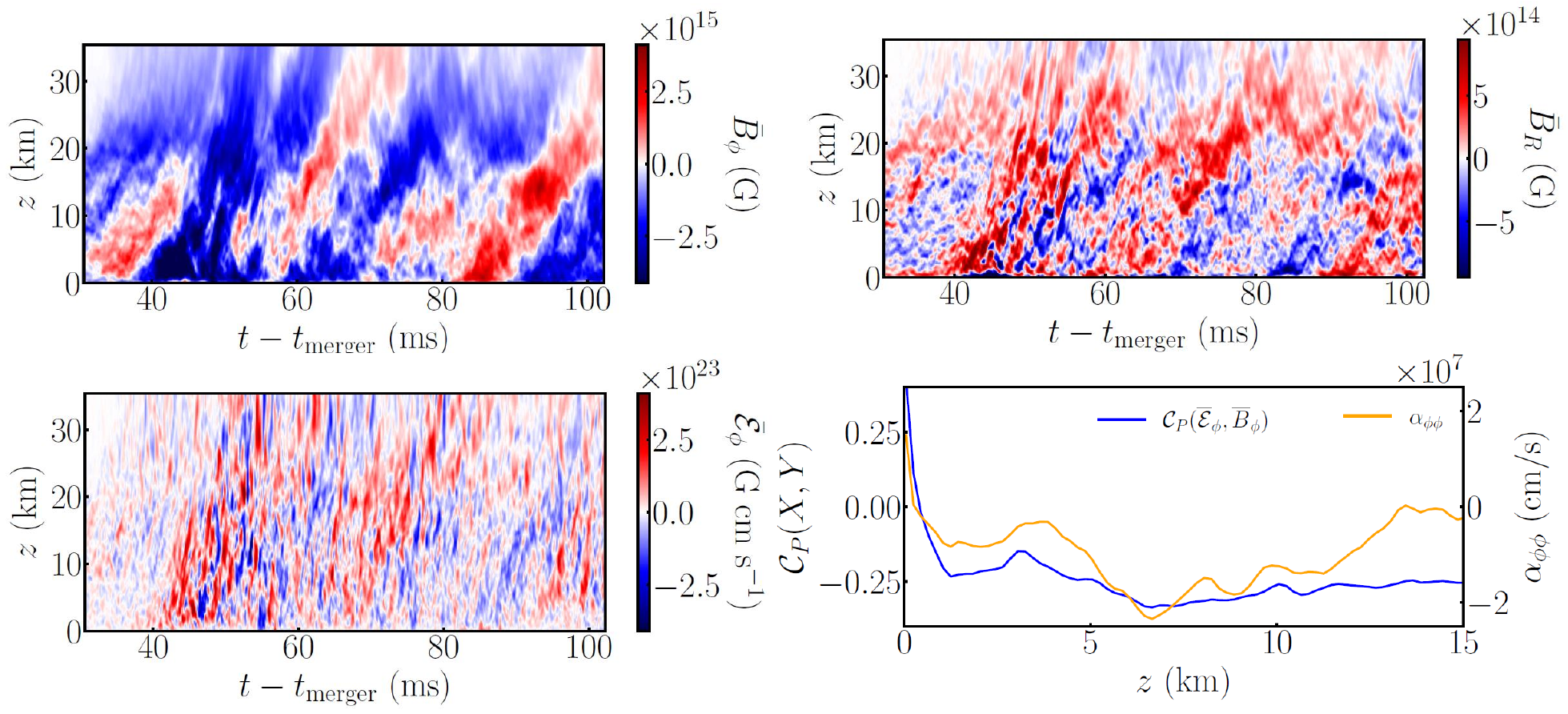}
\caption{\textbf{$\alpha\Omega$ dynamo inside the binary neutron star merger remnant.} Butterfly diagram at $R = 30$ km: (Top-left) Mean toroidal magnetic field $\bar{B}_\phi$. (Top-right) Mean radial magnetic field $\bar{B}_R$. (Bottom-left) Toroidal electromotive force $\bar{\mathcal{E}}_\phi$. (Bottom-right) $\alpha_{\phi\phi}$ parameters (orange) and correlation between $\bar{\mathcal{E}}_\phi$ and $\bar{B}_\phi$ (blue).
}\label{fig:BF}
\end{figure}

\begin{figure}
\centering
\includegraphics[width=1.0\textwidth]{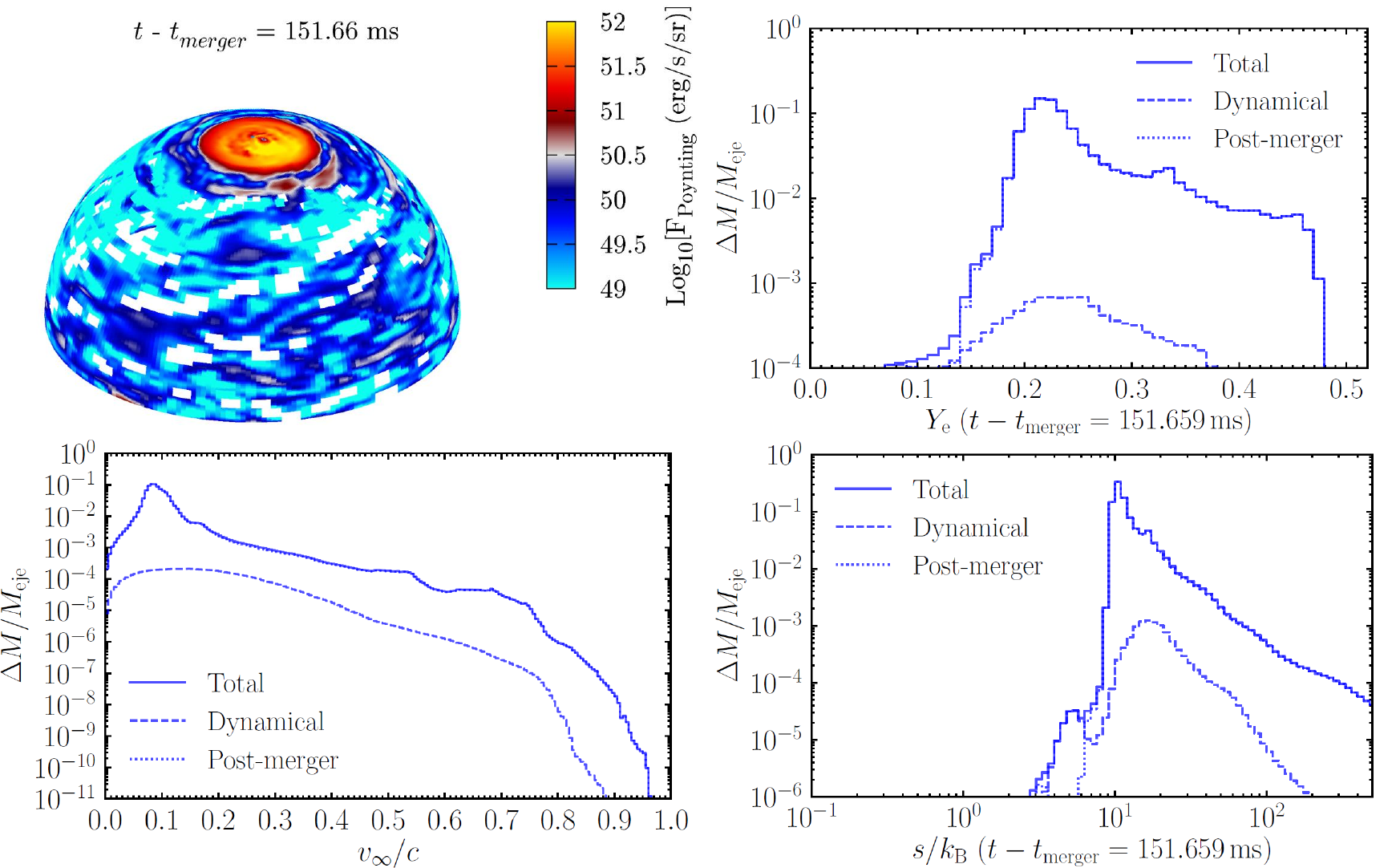}
\caption{\textbf{The electromagnetic signals from the binary neutron star merger remnants.}
 (Top-left) Angular distribution of the Poynting flux on a sphere with $r\approx 500$~km at $t-t_\mathrm{merger}\approx 150$~ms.
 (Top-right) The electron fraction distribution for the magnetically-driven post-merger ejecta (dashed) and the dynamical ejecta (dotted). (Bottom) Same as the top-right panel, but for the terminal velocity (left) and entropy per baryon (right).
}\label{fig:SGRB}
\end{figure}

\newpage

\begin{figure*}[t]
\centering
\includegraphics[width=0.6\textwidth]{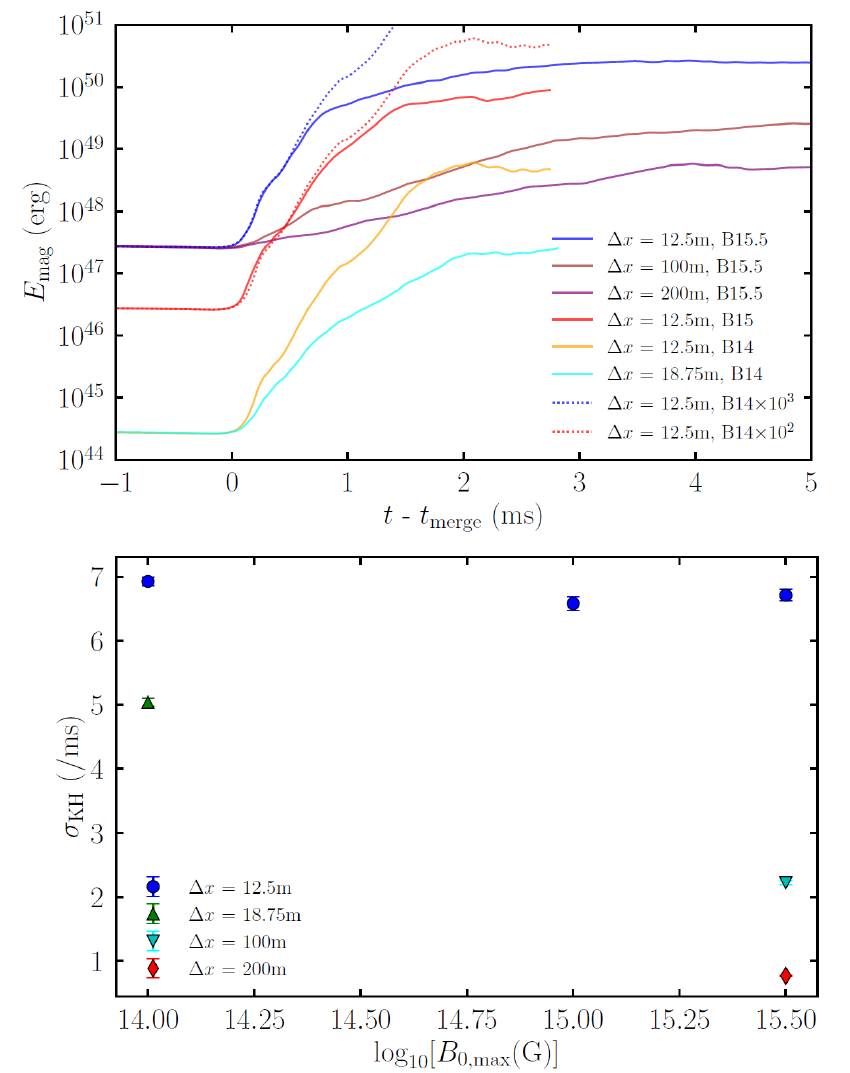}
\caption{{\bf Kelvin-Helmholtz instability growth rate at the merger.} (Top) Same as the inset in Fig.~1 in the main article, but with $\Delta x_{\rm finest} = 18.75$~m for $B_{0,\mathrm{max}}=10^{14}$~G (cyan). The blue- and red-dotted curves show the evolution for $\Delta x_{\rm finest}=12.5$~m and $B_{0,\mathrm{max}}=10^{14}$~G magnified by a factor of $10^3$ and $10^2$, respectively. (Bottom) Dependence of the growth rate of the electromagnetic energy at the merger due to the Kelvin-Helmholtz instability on the initial magnetic field strength $B_{0,\mathrm{max}}$ and grid resolution. The error is due to the fitting.Data are presented as mean median values +/- SD.}\label{fig:SM_KH}
\end{figure*}

\begin{figure*}[t]
\centering
\includegraphics[width=0.3\textwidth]{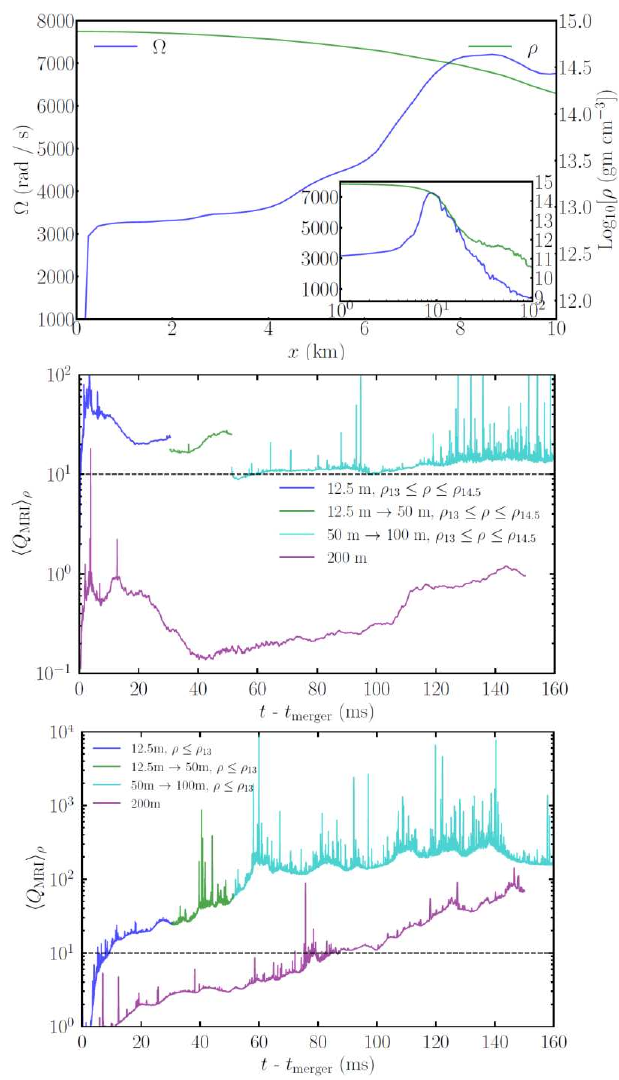}
\caption{{\bf Magnetorotational instability inside the merger remnant.} (Top) The radial profile of the angular velocity (blue) and the rest-mass density (green) on the equatorial plane at $t-t_\mathrm{merger}\approx 50$~ms. Magnetorotational instability is inactive in a region with $x \lesssim 9$~km and $\rho \gtrsim 10^{14.5}~\rm{g/cm^3}$. The inset shows the same profiles with the logarithmic scale in the radial direction. (Middle) Magnetorotational instability quality factor in a core region as a function of time. The remnant core is defined by a region with $\rho\ge 10^{13}~\rm{g/cm^{3}}$. The blue curve denotes the employed finest grid resolution of $12.5$~m. At $t-t_\mathrm{merger}\approx 30$~ms, the two finest domains with $\Delta x_{\rm finest}=12.5$~m and $\Delta x=25$~m are removed. Thus, the employed grid resolution is $\Delta x_{\rm finest}=50$~m plotted with the green curve. At $t-t_\mathrm{merger}\approx 50$~ms, the finest domain with $\Delta x_{\rm finest}=50$~m is removed and the subsequent evolution with $\Delta x_{\rm finest}=100$~m is plotted with the cyan curve. The result with $\Delta x_{\rm finest}=200$~m is plotted with the purple curve. (Bottom) The same as the middle panel, but for a torus defined by a region with $10^7~{\rm g/cm^3} \le \rho \le 10^{13}~{\rm g/cm^{3}}$. }\label{fig:SM_MRI}
\end{figure*}

\begin{figure*}[t]
\centering
\includegraphics[width=0.8\textwidth]{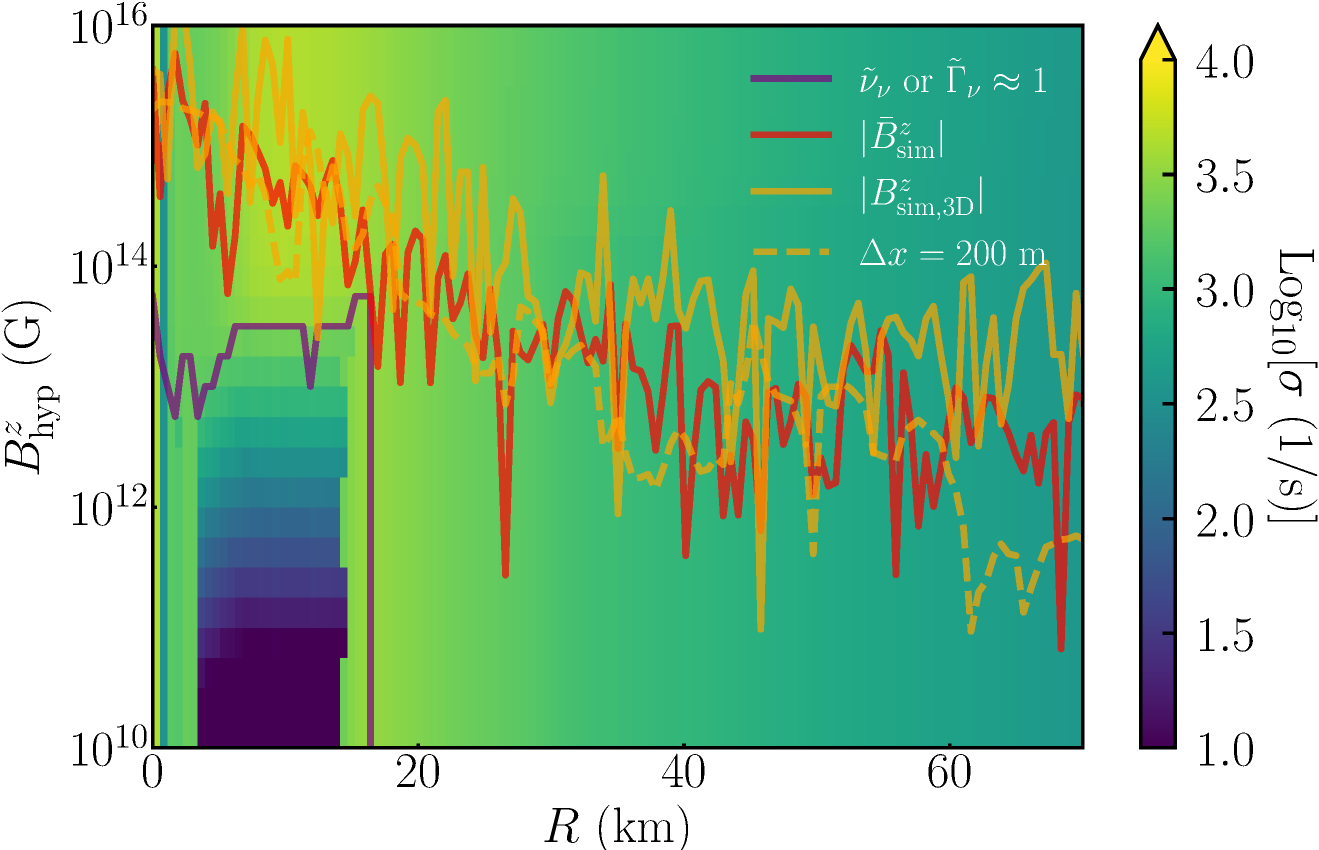}
\caption{{\bf Neutrino viscosity and drag on the magnetorotational instability.} Growth rate of the fastest growing mode of the neutrino viscous/drag magnetorotational instability as a function of the radius of the remnant massive neutron star and the hypothetical value of the $z$-component of the magnetic field $B^z_\mathrm{hyp}$. We take the simulation data on the orbital plane at $t-t_\mathrm{merger}\approx 10$~ms. The purple curve denotes the boundary where $\tilde{\nu}_\nu$ or $\tilde{\Gamma}_\nu \approx 1$. Outside the boundary, the growth rate is essentially the same as the ideal magnetorotational instability. The red curve denotes the $z$-component of the azimuthally averaged magnetic field strength in the simulation. The orange-solid (dashed) curve denotes the three-dimensional data for the $z$-component on the $x=0$ axis in the simulation with $\Delta x_\mathrm{finest}=12.5~(200)$~m.}\label{fig:SM_MRI_growth}
\end{figure*}

\begin{figure*}[t]
\centering
\includegraphics[width=0.8\textwidth]{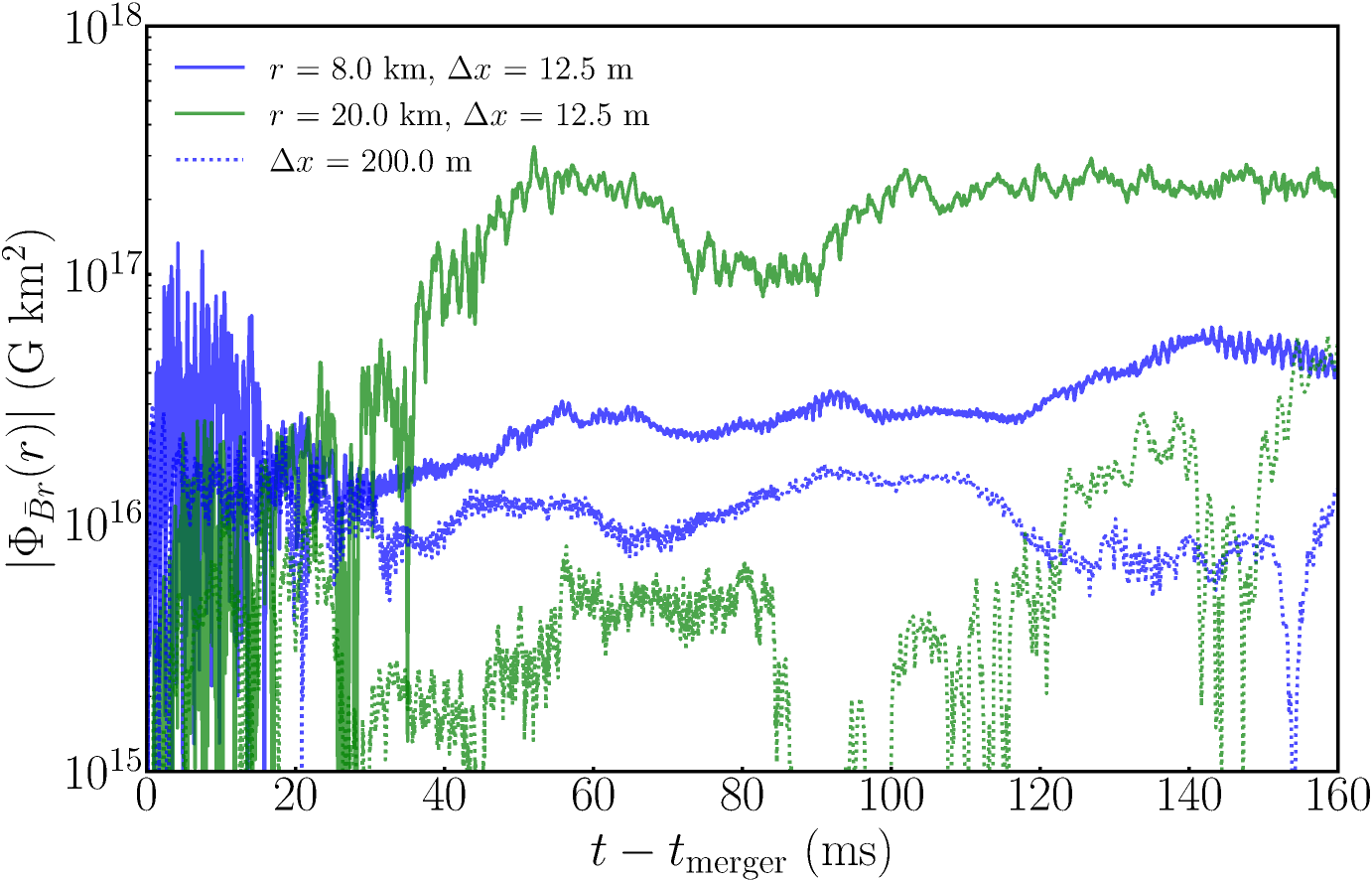}
\caption{{\bf Generation of the mean poloidal-magnetic flux inside the merger remnant.} Mean poloidal-magnetic flux as a function of the time after the merger. Blue and green-solid curves denote the flux on the sphere of $r=8$ and $20$~km, respectively, which are representative for the magnetorotational instability inactive and active region, in the simulation with $\Delta x_\mathrm{finest}=12.5$~m. The dashed counterparts denote the simulation with $\Delta x_\mathrm{finest}=200$~m. }\label{fig:SM_Mean_flux}
\end{figure*}

\begin{figure}[h!]
\centering
\includegraphics[width=1.0\textwidth]{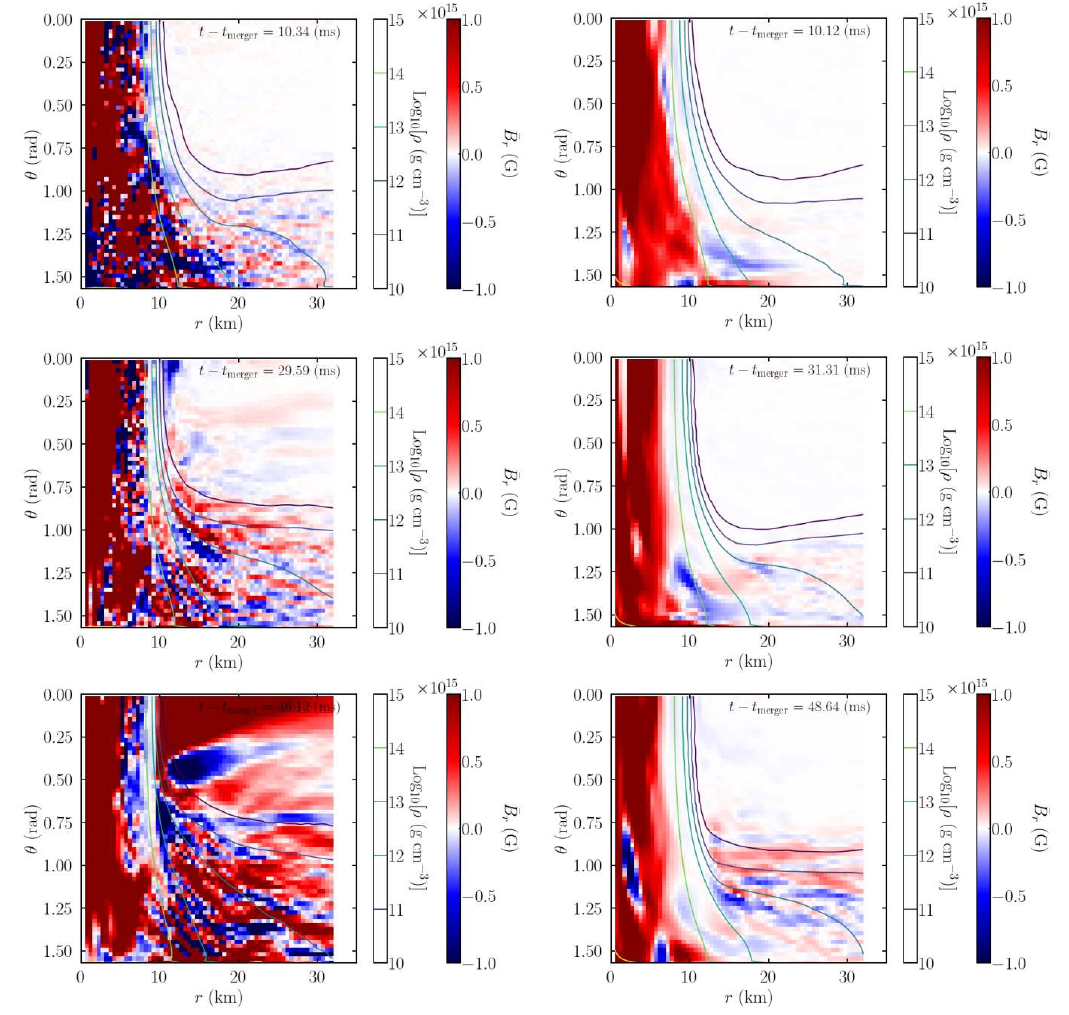}
\caption{{\bf Generation of the mean radial magnetic field.} Meridional profile of the mean radial magnetic field, $\bar{B}_r$, at $\approx 10$~ms (top), $30$~ms (center), and $50$~ms (bottom) for the simulation with $\Delta x_\mathrm{finest}=12.5$~m (left column) and $200$~m (right column). The rest-mass density contour curves are also plotted. The visualization is the following link (\url{http://www2.yukawa.kyoto-u.ac.jp/~kenta.kiuchi/anime/SAKURA/movie_Mean_Poloidal_Flux.mp4}) for the super-high resolution simulatoin, and \url{http://www2.yukawa.kyoto-u.ac.jp/~kenta.kiuchi/anime/SAKURA/movie_Mean_Poloidal_Flux_Low.mp4} for the low-resolution simulation.}\label{fig:SM_Mean_flux_prof}
\end{figure}

\begin{figure}[h!]
\centering
\includegraphics[width=0.64\textwidth]{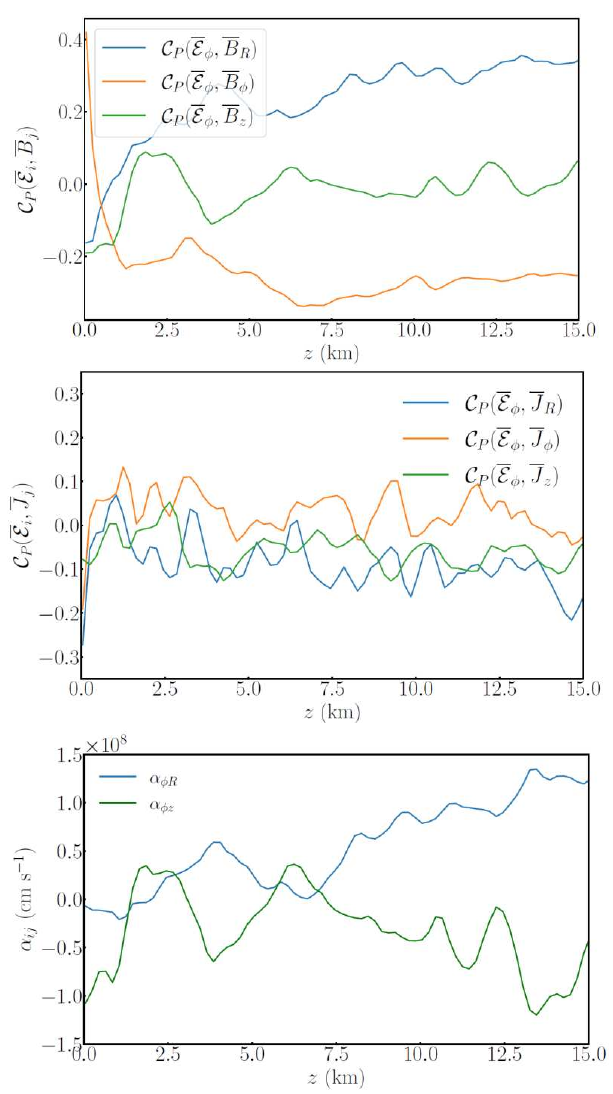}
\caption{{\bf $\alpha$ effect vs turbulent resistivity in the $\alpha\Omega$ dynamo.} (Top) Time-averaged correlations between the toroidal electromotive force $\bar{{\cal E}}_\phi$ and the mean magnetic field components $\bar{B}_j$ for $R = 30$ km. (Middle) Time-averaged correlations between the toroidal electromotive force $\bar{{\cal E}}_\phi$ and the mean current $\bar{J}_j$ components for $R = 30$ km. (Bottom) Alpha tensor components $\alpha_{\phi i}$ with $i \in [R, z]$ for $R = 30$ km.}\label{fig:emf_b_corr_SM}
\end{figure}

\begin{figure}[h!]
\centering
\includegraphics[width=1.0\textwidth]{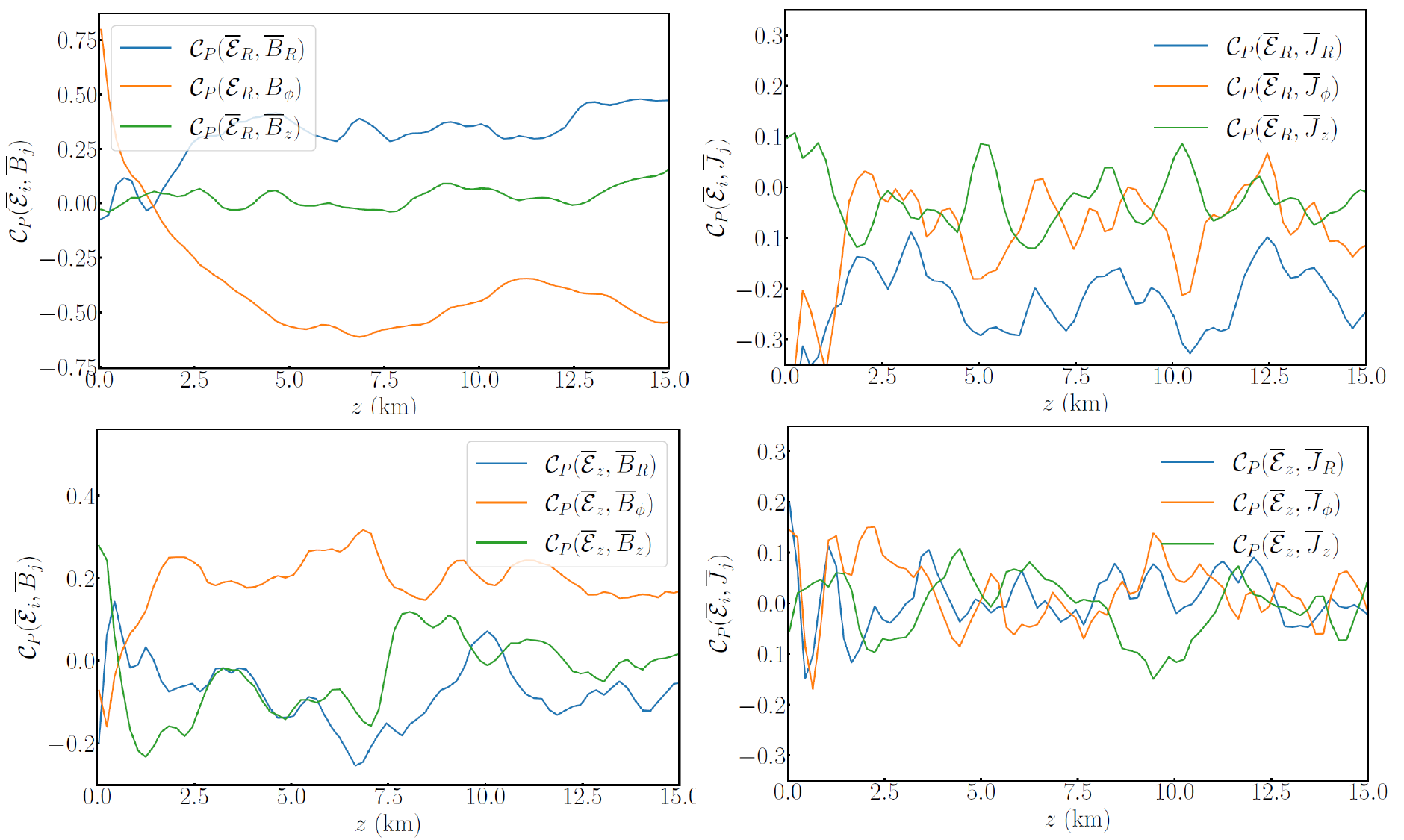}
\caption{{\bf $\alpha\Omega$ vs $\alpha^2\Omega$ dynamo.} (Left) Time-averaged correlations between the poloidal electromotive force $\bar{{\cal E}}_R$ (top) and $\bar{{\cal E}}_z$ (bottom) and mean magnetic field components $\bar{B}_i$ for $R = 30$ km. (Right) Time-averaged correlations between the poloidal electromotive force $\bar{{\cal E}}_R$ (top) and $\bar{{\cal E}}_z$ (bottom) and mean current components $\bar{J}_i$ for $R = 30$ km. }\label{fig:corr_EMF_r_z}
\end{figure}

\begin{figure}[h!]
\centering
\includegraphics[width=0.3\textwidth]{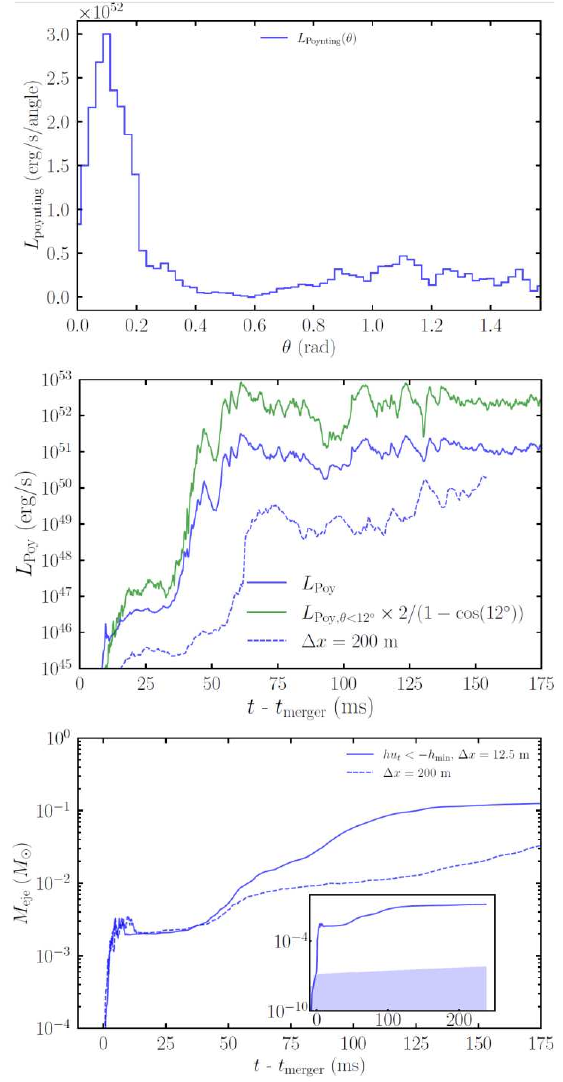}
\caption{{\bf Properties of the Poynting flux-dominated outflow and post-merger ejecta.} (Top) Angular distribution of the luminosity of the Poyinting flux at the end of the simulation of $t-t_\mathrm{merger}\approx 150$~ms. Angular distribution of the terminal Lorenz factor for the incipient magnetic-tower jet (blue) and the luminosity of the Poyinting flux (green) at the end of the simulation of $t-t_\mathrm{merger}\approx 130$~ms. (Middle) Luminosity for the Poynting flux as a function of the post-merger time. The green curve is the jet-opening-angle corrected luminosity. The blue-dashed curve plots the luminosity for the simulation with $\Delta x_{\rm finest}=200$~m. (Bottom) Ejecta as a function of the post-merger time. The solid curve denotes the ejecta satisfying the Bernoulli criterion. The colored region in the inset shows the violation of the baryon mass conservation.The blue-dashed curve plots the ejecta for the simulation with $\Delta x_{\rm finest}=200$~m.}\label{fig:Poynting_flux_SM}
\end{figure}





\end{document}